\documentclass[useAMS,usenatbib]{mn2e}
\newcommand{\beq}{\begin{equation}}
\newcommand{\eeq}{\end{equation}}

\newcommand{\apj}{ApJ}
\newcommand{\apjl}{ApJL}
\newcommand{\apjs}{ApJS}

\newcommand{\mnras}{MNRAS}
\newcommand{\aap}{A\&A}

\newdimen\hssize
\newdimen\hdsize
\usepackage{amssymb} 

\def\apj{\rm {ApJ}}                
\def\apjl{\rm{ApJL}}                
\def\apjs{\rm{ApJS}}               
\def\aap{\rm{A\&A}}                
\def\mnras{\rm{MNRAS}}             

\newcommand{\OIII}{[\rm O~ \sc{III}]}
\newcommand{\NII}{[\rm N~ \sc{II}]}

\usepackage{graphicx}

\title[The evolution and dependence of the local MZ relation] {The evolution and dependence of the local mass-metallicity relation}
\author[Wu et  al.]{Yu-Zhong Wu$^{}$\thanks{E-mail: yzwu@nao.cas.cn},
       Wei Zhang$^{}$,
       Yong-Heng Zhao$^{}$\\
       ${}$  Key Laboratory of Optical Astronomy, National Astronomical Observatories, Chinese Academy of Sciences, Beijing 100012, China\\}

\begin{document}

\date{Accepted ........ Received ........; in original form ........}
\pagerange{\pageref{firstpage}--\pageref{lastpage}} \pubyear{2010}
\maketitle
\label{firstpage}

\begin{abstract}

We present a sample of 86,111 star-forming galaxies (SFGs) 
selected from the catalogue of the MPA-JHU emission-line measurements for
SDSS DR7 to investigate the evolution of mass-metallicity (MZ) relation. 
We find that: under the $\rm log(L_{H \alpha})>41.0$ threshold, the
$0.04<z\leqslant0.06$ SFGs with $9.2<$ log($M_{\star}/M_{\sun})<9.5$ 
have always higher metallicities ($\sim 0.1$ dex) than the $0.10<z<0.12$ SFGs 
by using the closely-matched control sample method; under the 
$\rm log(L_{\OIII})>39.7$ threshold, the $0.04<z\leqslant0.06$ 
SFGs with $9.2<$ log($M_{\star}/M_{\sun})<9.5$ do not exhibit the 
evolution of the MZ relation, in contrast to the $0.10<z<0.12$ SFGs.
We find that the metallicity tends to be lower in galaxies
with a higher concentration, higher S\'{e}rsic index, or higher SFR.
In addition, we can see that the stellar mass and metallicity usually present
higher in galaxies with a higher $D_{n}4000$ or higher log(N/O) ratio.
Moreover, we present the two galaxy populations 
with log($M_{\star}/M_{\sun}$) below $10.0$ or greater than $10.5$ 
in the MZ relation, showing clearly an anticorrelation and a positive 
correlation between specific star formation rate and 12+log(O/H).

\end{abstract}


\section{INTRODUCTION}

The connection between stellar mass ($M_{\star}$) and gas-phase oxygen
abundance (metallicity, Z) is one of the most fundamental relations 
established in
the galaxy formation and evolution. The mass-metallicity (MZ) relation
was first shown by Lequeux et al. (1979), and later observations 
confirmed it with a larger sample from the Sloan Digital Sky
Survey (SDSS; Tremonti et al. 2004). Beyond the local universe, the
MZ relation has been observed to extend out to $z\sim3.5$, and
the metallicity decreases with increasing redshift at a given stellar
mass (Erb et al. 2006; Maiolino et al. 2008; Troncoso et al. 2014;
Sanders et al. 2015). With regard to the origin of the MZ relation,
physical explanations generally invoke an equilibrium between
metal-enriched gas outflows and pristine gas inflows (Mannucci et
al. 2010; Lara-L\'{o}pez et al. 2010; Brown et al. 2018), therefore
the change from the balance may be a tracer of understanding 
these mechanisms.

Some studies demonstrated the star formation rate (SFR) dependence of
MZ relations (Ellison et al. 2008; Mannucci et al. 2010; 
Lara-L\'{o}pez et al. 2010; Andrews \& Martini 2013). The dependence of 
the MZ relation on SFR originated from the work of Ellison et al. (2008), 
and Mannucci et
al. (2010) and Lara-L\'{o}pez et al. (2010) found almost simultaneously 
a tight relationship among SFR, stellar mass, and
metallicity for z$ \sim 0.1$ star-forming galaxies (SFGs), 
and concluded that the $\rm
M-\rm Z-SFR$ relation is invariant at $z<2.5$ and $z<3.5$, respectively. 
In addition, several studies reported that SFGs at $z\sim 1-3$ do not 
show any evidence for the redshift evolution of $\rm M-SFR-Z$ relation
(Christensen et al. 2012; Wuyts et al. 2012, 2016; Belli et al.
2013; Henry et al. 2013; Stott et al. 2013; Maier et al. 2014; Yabe
et al. 2015; Hunt et al. 2016; Sanders et al. 2018).

\begin{figure}
\begin{center}
\includegraphics[width=8cm,height=6cm]{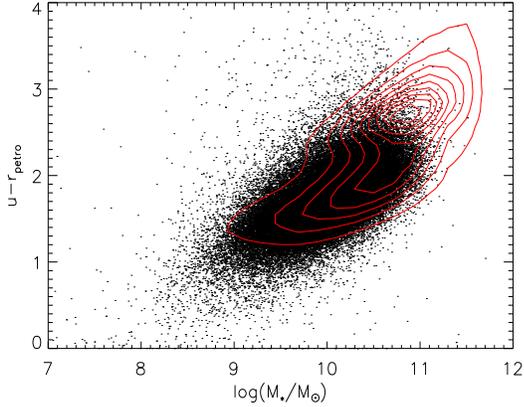}
\caption{The stellar mass and u-r colour diagram. Our SFG 
sample occupys the blue sequence (black dots), and the red contours 
represent the distribution of the whole SDSS DR7 sample.}
\end{center}
\end{figure}

Moreover, Salim et al. (2014) studied the relation between metallicity 
and specific star formation rate (sSFR=SFR/$M_{\star}$) in 
distinct mass bins by analyzing the SDSS data, and found a 
significant correlation yet fairly weaker than that shown by 
Mannucci et al. (2010). Collecting 1381 field galaxies at 
$0.5 \leqslant z \leqslant0.7$ from deep spectroscopic survey, 
Guo et al. (2016) found that the dependence of MZ relations on SFR 
is weaker than that presented in local galaxies.

Utilising the CALIFA data to study the MZ relation, S\'{a}nchez et al. 
(2013) did not find a correlation between the MZ relation and SFR.
Recently, Barrera-Ballesteros et al. (2017) presented the integrated
MZ relation with more than 1700 MaNGA SFGs, and found no dependence
of the MZ relation on SFR. Besides, many other literatures presented
that galaxies at $z\sim 1-3$ do not trace the predictions
of the $z\sim0$ FMR, indicating an evolving $\rm M-SFR-Z$
relation (Cullen et al. 2014; Yabe et al. 2014; Zahid et al. 2014;
Wuyts et al. 2014; Sanders et al. 2015, 2018; Salim et al. 2015;
Kashino et al. 2017).

Since many studies mentioned above focus on the SFR dependence of MZ
relations, we restrain the stellar mass and SFR to explore the
redshift evolution of the MZ relaition. In addition, we revisit the
evolution of the MZ relation and investigate the MZ relation
with the S\'{e}rsic index, concentration index, and
$D_{n}4000$. The paper is organized as follow. In Section 2, we show
the SFG sample and various data. We explore the evolution of the MZ
relation, and investigate the dependence and scatter of the MZ
relation with the S\'{e}rsic index, concentration index, SFR, 
sSFR, log(N/O), and $D_{n}4000$ in Section 3. In Section 4, the 
results and conclusions are summarized. We adopt 
$H_0=70~\rm km~s^{-1}~Mpc^{-1}$, $\Omega_{M}=0.3$, and 
$\Omega_{\Lambda}=0.7$.

\begin{figure*}
\begin{center}
\includegraphics[width=8cm,height=6cm]{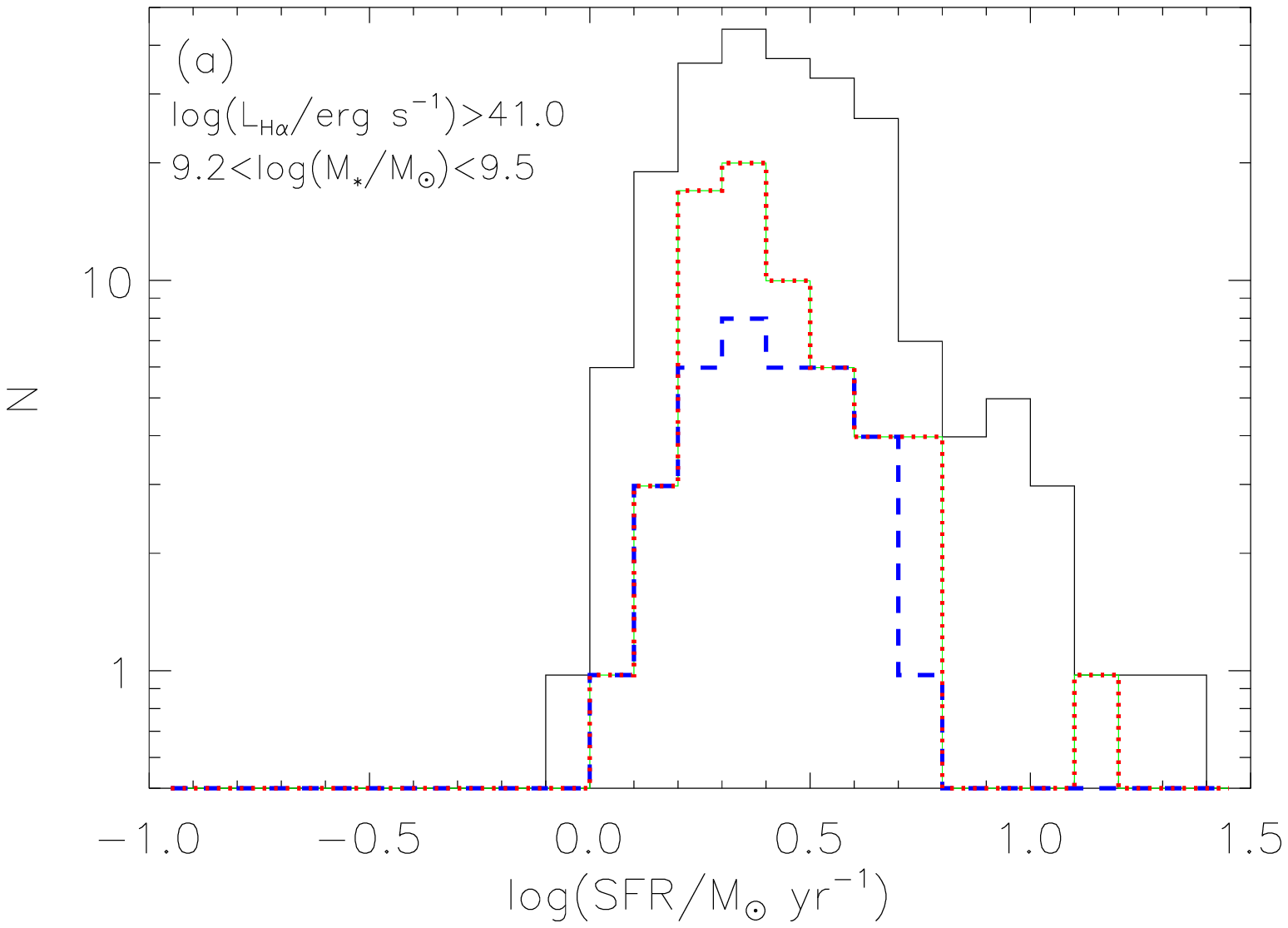}
\includegraphics[width=8cm,height=6cm]{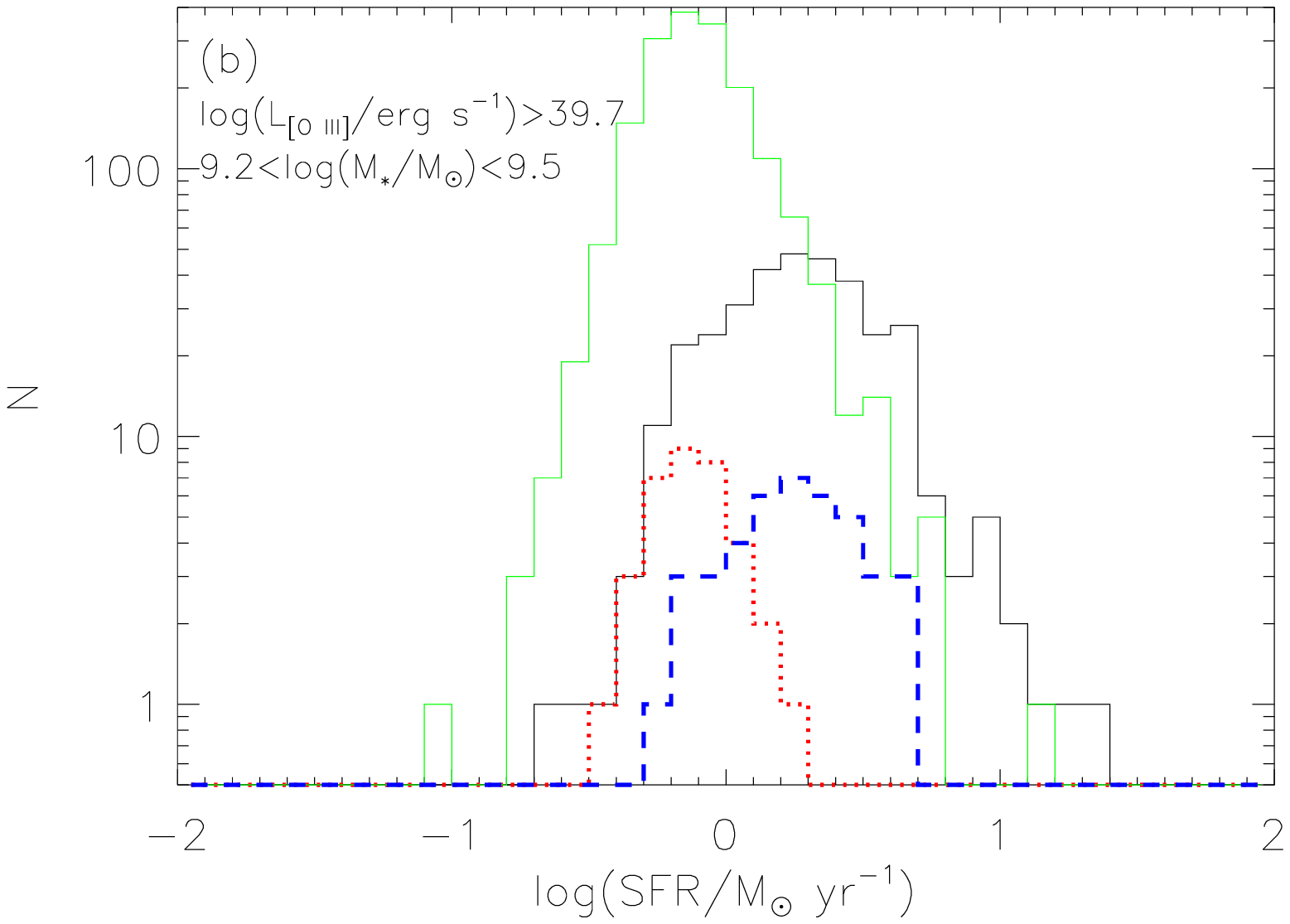}
\caption{Comparison of the SFR distributions of 
different sample SFGs
with $9.2<$log$(M_{\star}/M_{\sun})<9.5$. Panel(a) 
for log(L$_{\rm H\alpha})>41.0$.
Panel(b) for log(L$_{\rm [O III]})>39.7$. The green and black lines are
the SFR distributions of SFGs at 
$0.04<z\leqslant0.06$ and
$0.10<z<0.12$, respectively. The red dotted (blue dashed) lines
show the SFR distributions of control samples, which 
have the same SFR distribution as the SFGs at $0.04<z\leqslant0.06$
($0.10<z<0.12$), and come randomly from the SFGs at $0.10<z<0.12$ 
($0.04<z\leqslant0.06$), respectively.}
\end{center}
\end{figure*}

\begin{table*}
\caption{Summary of the evolution of the MZ relations for different
sample SFGs.}
\begin{scriptsize}
\begin{center}
\setlength{\tabcolsep}{2.5pt}
\renewcommand{\arraystretch}{1.2}
\begin{tabular}{cccccccccccccccl}
\hline \hline  & low/high z & CSs&\multicolumn{2}{c} {log(SFR)} &
$p_{\rm null}(\%)$& \multicolumn{2}{c} {log($M_{\star}$)} &$p_{\rm
null}(\%)$&
\multicolumn{2}{c} {$12+\rm log(O/H)$} & $p_{\rm null}(\%)$  \\
&&&\multicolumn{2}{c}{$M_{\sun}/yr^{-1}$}& &\multicolumn{2}{c}{$M_{\sun}$}&&&\\
\cline{4-5}\cline{7-8}\cline{10-11}
(1)& (2) & (3) & (4) &(5)&(6) &(7)&(8) &(9)&(10)&(11)&(12) \\
\hline
$\rm log (L_{H \alpha})>41.0$&66  &66&$0.36\pm0.19$&$0.37\pm0.19$&54&$9.40\pm0.09$&$9.41\pm0.08$&83&$8.71\pm0.11$&$8.57\pm0.12$&$4.6\times10^{-5}$&         \\
                             &225 &35&$0.42\pm0.32$&$0.37\pm0.17$&53&$9.42\pm0.08$&$9.40\pm0.09$&85&$8.60\pm0.13$&$8.71\pm0.09$&0.0296  &         \\
                             \hline

$\rm log(L_{\OIII})>39.7$    &1714&35&$-0.11\pm0.21$&$-0.15\pm0.15$&74&$9.35\pm0.09$&$9.39\pm0.09$&22&$8.76\pm0.10$&$8.78\pm0.11$&51  &    \\
                             &338 &41&$0.27\pm0.36$&$0.22\pm0.24$&65&$9.40\pm0.08$&$9.41\pm0.09$&68&$8.66\pm0.14$&$8.78\pm0.11$&0.00142 &       \\

\hline \hline
\end{tabular}
\parbox{6.5in}
{\baselineskip 11pt \noindent \vglue 0.5cm {\sc Note}: Cols.(1): The
different samples. Cols.(2) and (3) are the number of the samples of
$0.04<z\leqslant0.06$ (the lower redshift) or
$0.10<z<0.12$ (the higher redshift) and their corresponding control
samples(CSs). Cols.(4), (7), and (10): the
median value of log(SFR), log($M_{\star}$), and 12+log(O/H),
respectively, which comes from the lower (higher) redshift sample. 
Cols.(5), (8), and (11): the
median value of log(SFR), log($M_{\star}$), and 12+log(O/H),
respectively, which comes from their corresponding control samples (CSs).
Cols.(6), (9), and (12):
the probabilities for the null hypothesis that the two 
distributions are drawn randomly from the same population.}
\end{center}
\end{scriptsize}
\end{table*}

\section{THE DATA}

The sample in this study is selected from the catalogue 
of the SDSS Data Release 7 (DR7; Abazajian et al. 2009).
We \textbf{utilise} measurements of emission line fluxes, SFRs, and stellar 
masses, which are publicly available from the catalogue of 
Max Planck Institute for Astrophysics$-$John Hopkins University 
(MPA-JHU) SDSS DR7 release. To avoid the bias of the MZ relation from 
the aperture effect (Kewley, Jansen \& Geller 2005) and the selection 
effect of the metallicity evolution (Zahid et al. 2013; Wu et al. 2016), 
these galaxies are required to have the lower and
upper redshift limits of 0.04 and 0.12. Also, the covering fractions
are $>20\%$ of all galaxy fluxes, which are calculated from the r band
fiber and Petrosian magnitudes. According to the BPT diagram 
(Baldwin et al. 1981; Kauffmann et al. 2003a; Kewley et al. 2006), we 
obtain the SFGs from the above sample by using the equation 
(log([O~III]$/\rm H\beta)<0.61/[log(\NII/\rm H\alpha)-0.05]+1.3$) 
of Kauffmann et al. (2003a), forming an initial 
sample of 136,964 galaxies.

\begin{figure*}
\begin{center}
\includegraphics[width=14cm,height=10.5cm]{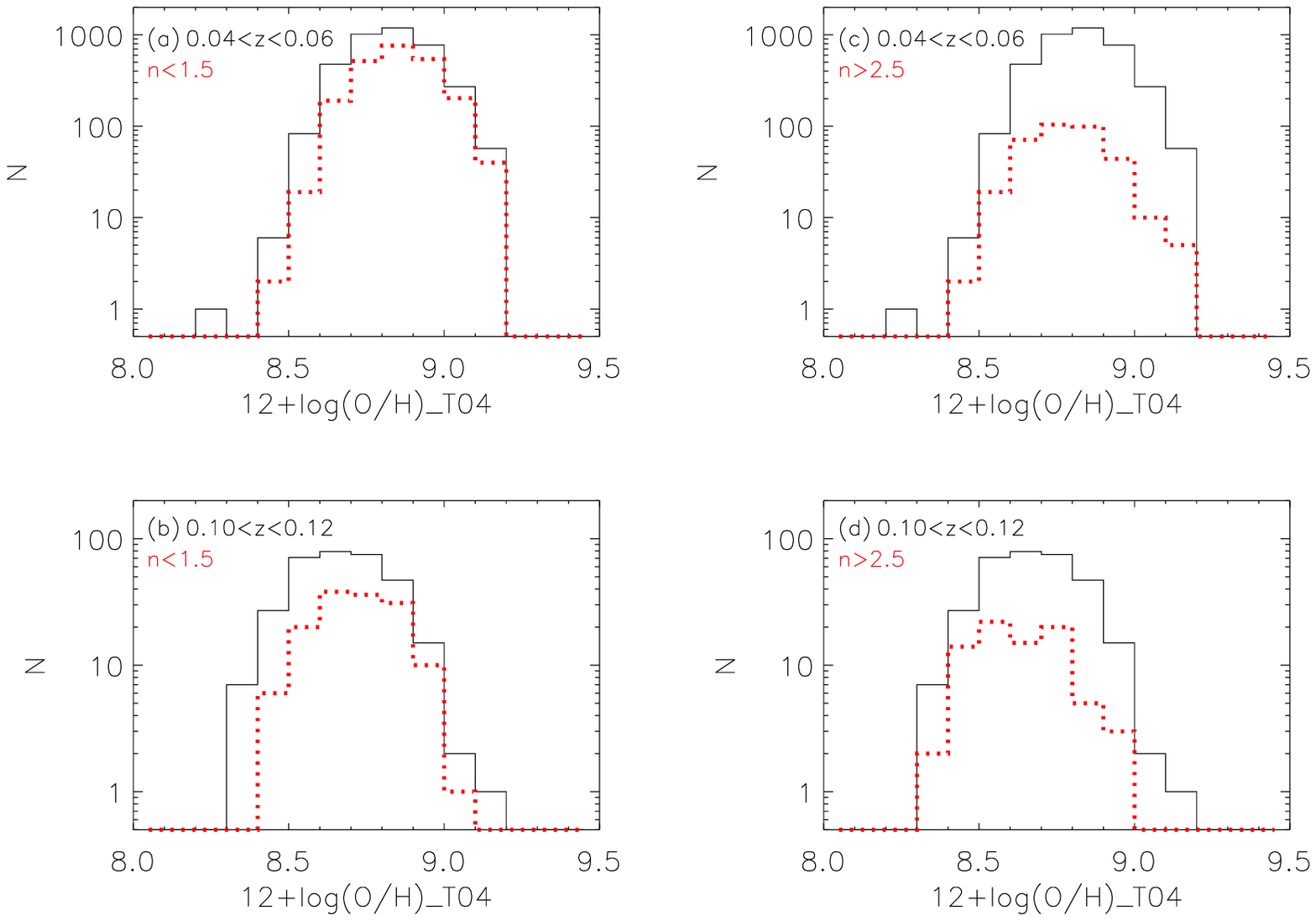}
\caption{Comparison of the metallicity distributions 
of different S\'{e}rsic index SFG samples
with $9.2<$log$(M_{\star}/M_{\sun})<9.5$. The black line 
of Fig. 3(a) (Fig. 3(b)) is the metallicity 
distribution of $0.04<z\leqslant0.06$ ($0.10<z<0.12$) SFGs with 
$9.2<$log$(M_{\star}/M_{\sun})<9.5$. Fig. 3(a) (Fig. 3(b)) 
has the same black line as Fig. 3(c) (Fig. 3(d)). 
The red dotted line of Fig. 3(a) (Fig. 3(c)) is the metallicity 
distribution of $0.4<z\leqslant0.06$ SFGs with
$n<1.5$ ($n>2.5$) and $9.2<$log$(M_{\star}/M_{\sun})<9.5$.
The red dotted line of Fig. 3(b) (Fig. 3(d)) is the metallicity 
distribution of $0.10<z<0.12$ SFGs with
$n<1.5$ ($n>2.5$) and $9.2<$log$(M_{\star}/M_{\sun})<9.5$.}
\end{center}
\end{figure*}

We use the emission line fluxes of [O~II]~$\lambda \lambda 3727,
3729$, H$\beta$, [O~III]~$\lambda \lambda 4959, 5007$, 
H$\alpha$, and [N~II]~ $\lambda 6584$ to analyze the MZ relation. The 
signal-to-noise ratio (S/N) cut of [O~III]~$\lambda 5007$ may bias the MZ 
relation, due to that this emission line often appears in high metallicity 
galaxies (Foster et al. 2012). Therefore, we choose the SFGs with 
S/N $>3$ for H$\beta$, H$\alpha$, [O~II]~$\lambda \lambda 3727,3729$, 
and [N~II]~ $\lambda 6584$. Since an SFR FLAG keyword shows 
the status of the SFR measurements, the keyword of $0$ is required. 
To break the degeneracy between the upper and lower branch 
$R_{23}$ solutions, we choose the SFGs with 
log([N~II] $\lambda6584$/[O~II]$\lambda3727$)$>-1.2$ in our 
sample (Kewley \& Ellison 2008). In fact, we find that the N2O2 ratio
excludes only one SFG from the above sample, so it does not introduce 
a selection bias to our final sample. Finally, we obtain the final 
sample of 86,111 SFGs.

Regarding SFRs and $M_{\star}$, we use a Chabrier (2003) initial mass
function (IMF) to correct them assumed a Kroupa (2001) IMF by dividing 
it by 1.06. To study the MZ relation, we use the
Petrosian half-light radii $R_{50}$ and $R_{90}$, which are the
radius enclosing $50\%$ and $90\%$ of the Petrosian flux,
respectively (Shen et al. 2003), and are from the New York
University Value-Added Galaxy Catalogue (NYU-VAGC; Blanton et al.
2005). With regard to the Petrosian flux, please see Section 2.1 of
Shen et al. (2003). Provided the radii $R_{50}$ and $R_{90}$, the
concentration index of a galaxy is defined as $c=R_{90}$/$R_{50}$,
which is related to galaxy morphological types. Also, we introduce
a $D_{n}4000$ index to study the MZ relation. The index is 
defined as a flux ratio in two spectral windows, which is
close to the 4000{\AA} break (Balogh et al. 1999): 3850-3950{\AA}
and 4000-4100{\AA} (Zahid \& Geller 2017). The index increases with
the stellar population age (Kauffmann et al. 2003a; Zahid et al.
2015). In addition, we adopt $n$ to represent the S\'{e}rsic
index, obtained from the NYU group (Zahid \& Geller 2017). The
S\'{e}rsic index ($n$) ranges are from $0$ to $5.9$. For
quiescent galaxies, Zahid \& Geller (2017) showed that galaxies with
higher $D_{n}4000$ often have larger S\'{e}rsic indexes
at a given stellar mass. Due to the two indexes(c and n) estimated 
from the NYU-VAGC, we cross-match it with the MPA-JHU SDSS DR7
catalogue within $2''$, and obtain 81,125 SFGs. In this paper, we will 
utilise a sample of 86,111 SFGs if the two indexes are not involved 
in our studies.

In Figure 1, we present the diagram of stellar mass 
and u-r colour (from Petrosian magnitudes). The red 
contours of Figure 1 are the 
distribution of 862,079 galaxy sample, which comes from 
the cross-match
of the NYU-VAGC and MPA-JHU SDSS DR7 catalogues. 
Our sample of 86,111 SFGs is shown with the black dots in Figure 1.
From Figure 1, we find that all SFGs 
of our sample are almost the late-type galaxy.
For the morphologies of S\'{e}rsic indexes, 
we use $n>2.5$ and $n<1.5$ to represent the spheroid-dominated 
and disk-dominated 
galaxies, respectively (Maier 2009). The concentration indexes 
$c\geqslant2.86$ and $c<2.86$ are used to separate early-type and 
late-type galaxies (Nakamura et al. 2003; Shen et al. 2003). 
In addtion, we show that these galaxies with $n>2.5$ 
or $c>2.86$ have lower matallicities compared to the late-type 
galaxies in Figures 3(c) and 3(d) or Figures 4(c) and 4(d). 
This is not consistent with that early-type galaxies generally
have higher stellar metallicity than the late-type galaxies
with larger difference in lower masses (Peng et al. 2015). 
In fact, these galaxies are likely to be
compact SFGs which usually show lower gas metallcity
than the normal SFGs (Hoopes et al. 2007).
In this paper, the two indexes (c and n) only represent
the galaxy structure properties.

In this paper, we use the $R_{23}$ method to estimate oxygen
abundances of SFGs (Pilyugin et al. 2006, 2010; Wu \& Zhang 2013),
and adopt the calibration of T04. In addition to the T04 method, 
we use the five methods of
metallicity estimators, for instance, Dopita 2016 (D16), 
Pettini \& Pagel 2004 (PP04-O3N2 and PP04-N2), Zaritsky, Kennicutt
\& Huchra 1994 (Z94), and Sanders et al. 2018 (Sander18).

\section{Results}

In this section, we first explore the evolution of the MZ relation 
using a control sample method under the emission line luminosity 
limits for log(H$\alpha$) and log($\OIII$) and using comparison 
of metallicity differences under the different galaxy morphologies.
Then we investigate the dependence of MZ relations of
$0.04<z\leqslant0.06$ and $0.10<z<0.12$ SFGs on the
S\'{e}rsic index, concentration index, SFR, sSFR, 
log(N/O), and $D_{n}4000$. Finally, we study the scatter of MZ 
relations with different metallicity estimators, 
S\'{e}rsic index, and concentration index.

\begin{figure*}
\begin{center}
\includegraphics[width=14cm,height=10.5cm]{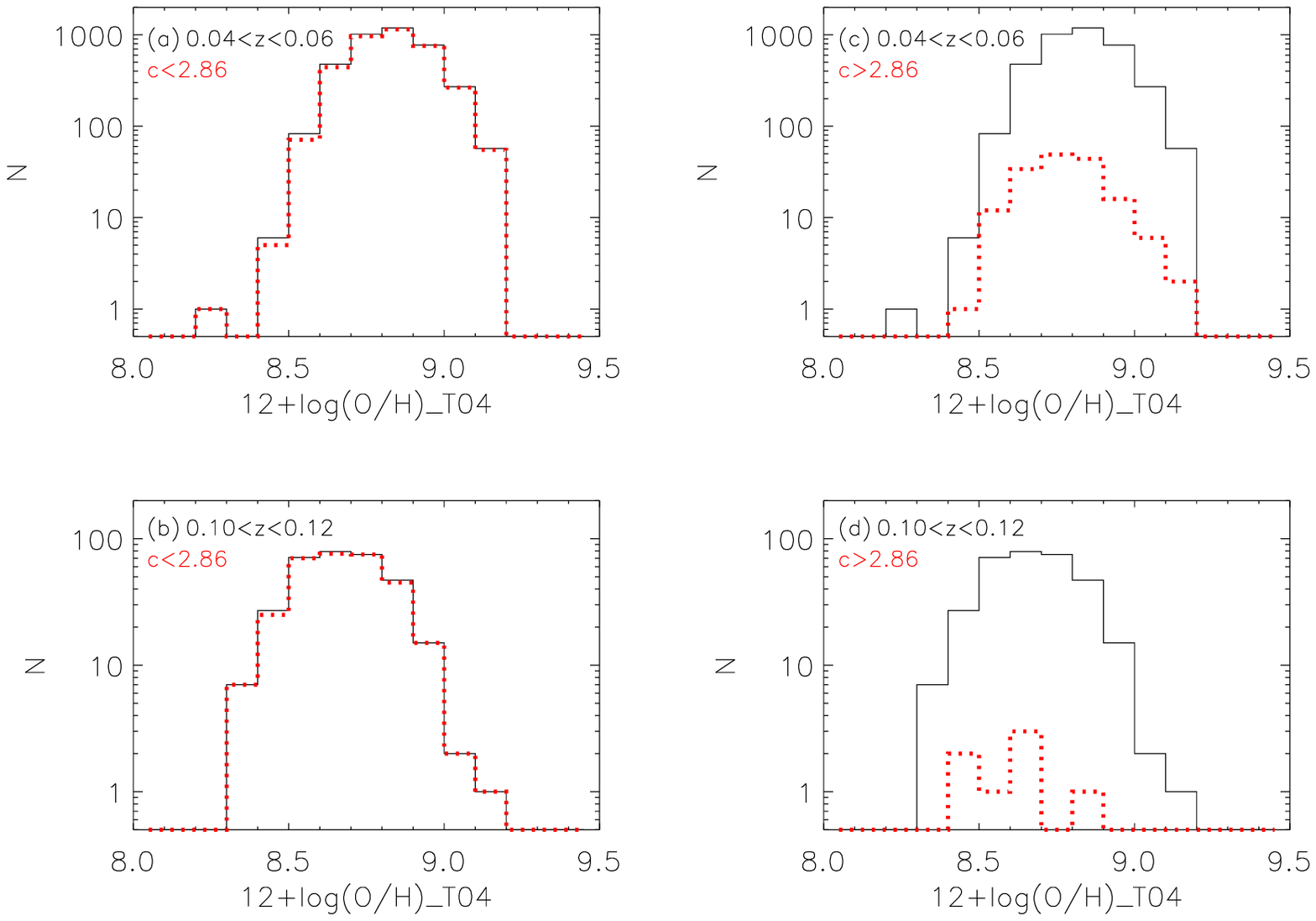}
\caption{Comparison of the metallicity distributions
of different concentration index SFG samples
with $9.2<$log$(M_{\star}/M_{\sun})<9.5$. 
The black line of Fig. 4(a) (Fig. 4(b)) is the metallicity 
distribution of $0.04<z\leqslant0.06$ ($0.10<z<0.12$) SFGs with 
$9.2<$log$(M_{\star}/M_{\sun})<9.5$. Fig. 4(a) (Fig. 4(b)) 
has the same black line as Fig. 4(c) (Fig. 4(d)). The red dotted 
line of Fig. 4(a) (Fig. 4(c)) is the metallicity 
distribution of $0.4<z\leqslant0.06$ SFGs with
$c<2.86$ ($c\geqslant2.86$) and $9.2<$log$(M_{\star}/M_{\sun})<9.5$.
The red dotted line of Fig. 4(b) (Fig. 4(d)) is the metallicity 
distribution of $0.10<z<0.12$ SFGs with
$c<2.86$ ($c\geqslant2.86$) and $9.2<$log$(M_{\star}/M_{\sun})<9.5$.}

\end{center}
\end{figure*}

\subsection{The evolution of the MZ relation}

In the SDSS sample, Juneau et al. (2014) found \textbf{an} evolution
 of log($L_{\rm H \alpha}$) and $\rm log(L_{\OIII}$). They got 
the observed limits of log($L_{\rm H \alpha}$) and $\rm log(L_{\OIII}$), and
presented an artificial evolution of the MZ relation. 
In the $0.04<z<0.12$ SDSS sample, Wu et al. (2016) also investigated 
and confirmed \textbf{the} redshift evolution of log($L_{\rm H \alpha}$), and 
$\rm log(L_{\OIII}$). Wu et al.
(2016) used the MZ relation of the $0.10<z<0.12$ subsample to
correct the artificial evolution of MZ relations, and 
obtained the minimum luminosity limits
of $\rm log (L_{H \alpha})>41.0$ and $\rm log(L_{\OIII})>39.7$. 
In this article, we use a new method, 
the closely-matched control sample (Zhang, Kong \& Cheng 2008), to 
check the evolution of the MZ relation. The control sample is constructed 
from the corresponding sample, which is matched in $M_{\star}$ and SFR. 
For instance, regarding the $0.04<z\leqslant0.06$ SFG sample and its 
corresponding sample ($0.10<z<0.12$ SFGs), the control sample 
has the same SFR distribution as the SFGs at $0.04<z\leqslant0.06$, and 
is randomly from the $0.10<z<0.12$ SFGs. In Wu et al. (2016), 
the metallicity difference of $\sim0.15$ 
dex in the MZ relations happened 
at log($M_{\star}/M_{\sun})\sim 9.3$, so we choose 
$9.2<$ log$(M_{\star}/M_{\sun})<9.5$ as the stellar mass 
range matched. With 
regard to SFR matched, we employ the low (high) redshift range SFG sample and 
its corresponding control sample having the same SFR distribution. For
comparision, all samples have the same stellar mass range matched of
$9.2<$ log$(M_{\star}/M_{\sun})<9.5$.

In Figure 2(a), comparison of all samples at the two redshift ranges 
are based on the luminosity limit of $\rm log (L_{H \alpha})>41.0$.
The green and black lines represent the distributions 
of SFRs of SFGs at $0.04<z\leqslant0.06$ and $0.10<z<0.12$, respectively. 
The red dotted line shows the distribution of SFRs of the control
sample, which has the same SFR distribution as the
$0.04<z\leqslant0.06$ SFGs, and comes randomely from the SFGs at
$0.10<z<0.12$. The blue dashed line represents the
distribution of SFRs of the control sample, which 
has the same SFR distribution as the $0.10<z<0.12$ SFG sample, 
and comes randomly from the SFGs at $0.04<z\leqslant0.06$.

\begin{figure*}
\begin{center}
\includegraphics[width=16cm,height=12cm]{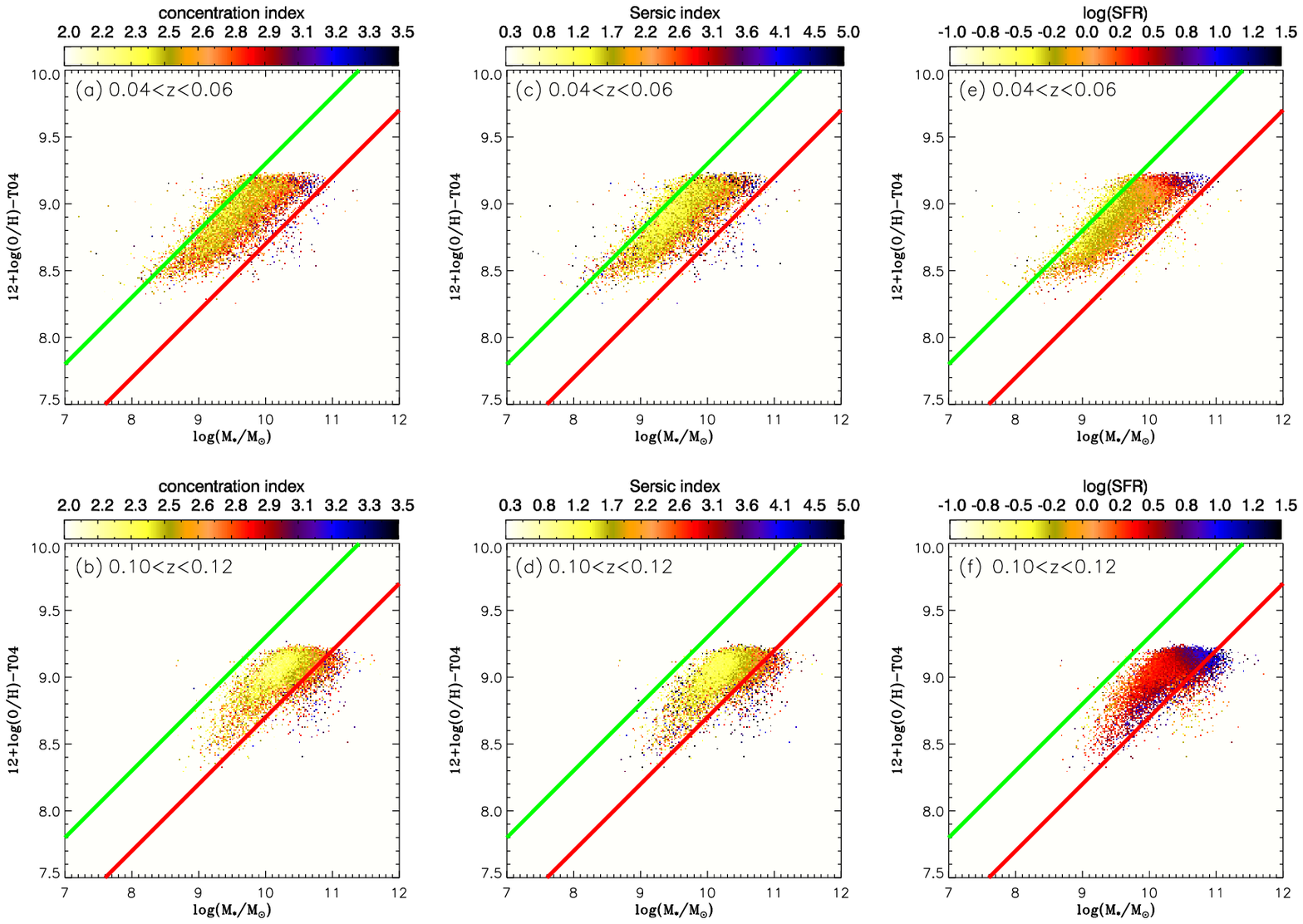}
\caption{Comparison of the MZ relations of $0.04<z\leqslant0.06$
and $0.10<z<0.12$ SFGs using the colour bars of the concentration
index and S\'{e}rsic index, and SFR. The green and red
lines are the diagnostic lines (is shown for reference) for 
comparing the MZ relations at different 
redshift ranges. The metallicity is estimated by the T04 method.}
\end{center}
\end{figure*}

\begin{figure*}
\begin{center}
\includegraphics[width=16cm,height=12cm]{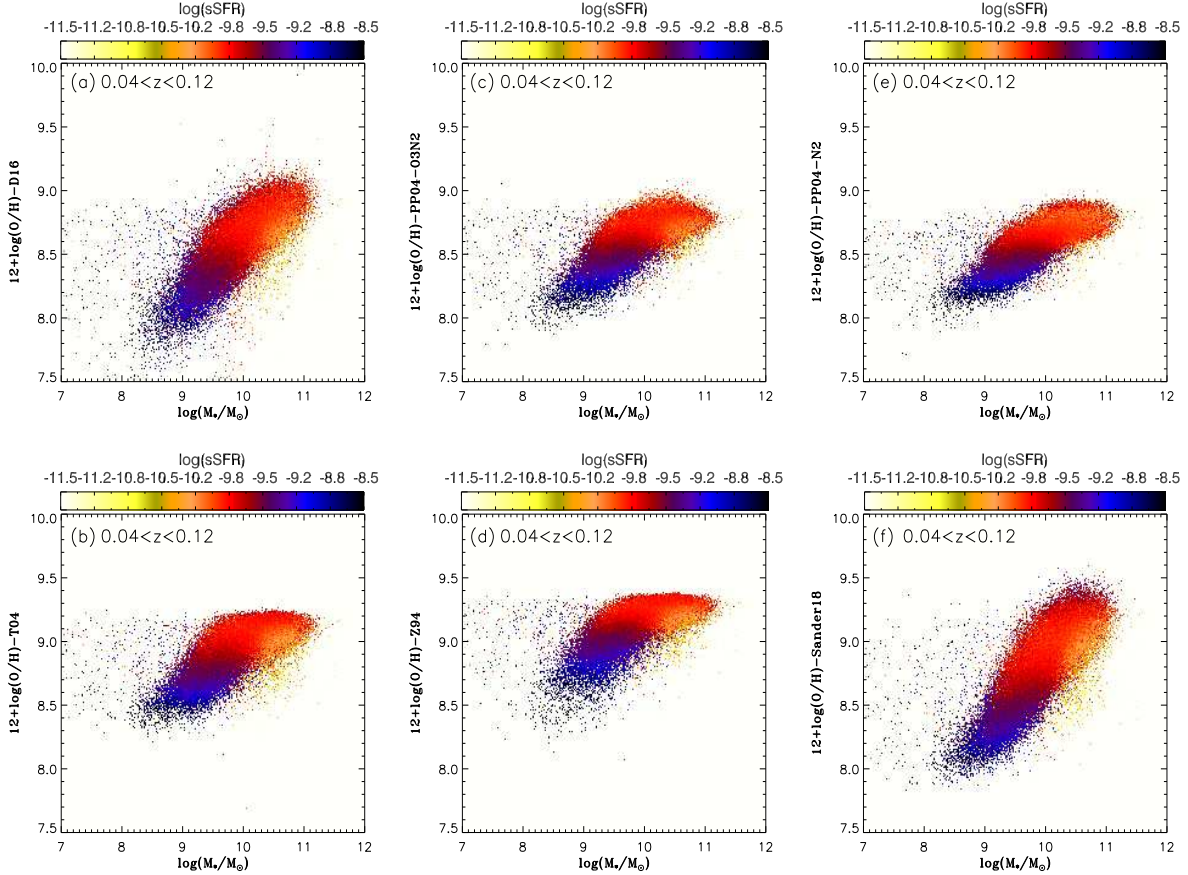}
\caption{Comparison of the MZ relations of $0.04<z<0.12$ SFGs 
with the colour bars of sSFR for different metallicity estimators.}
\end{center}
\end{figure*}

\begin{figure*}
\begin{center}
\includegraphics[width=14cm,height=10.5cm]{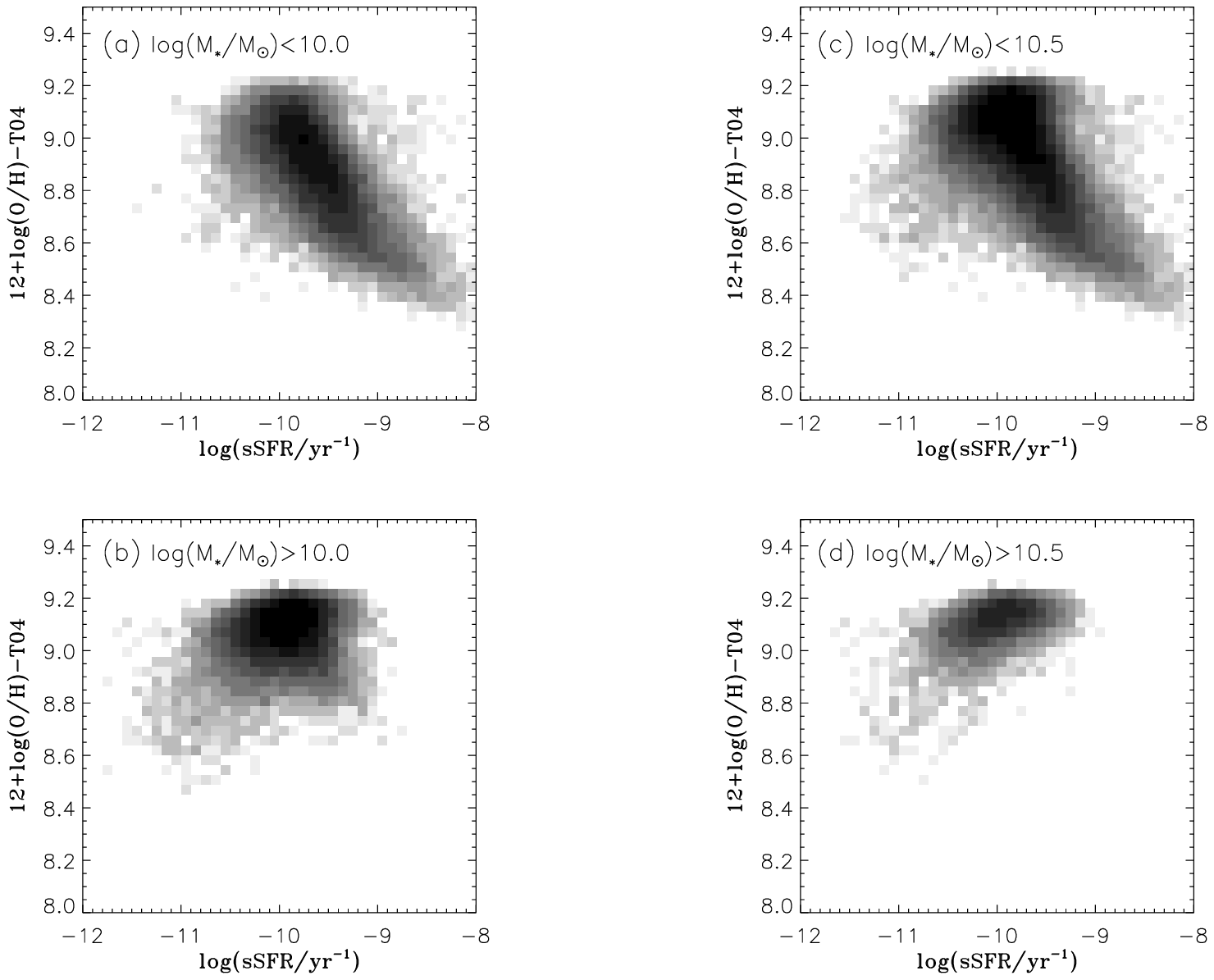}
\caption{Comparison of the correlations between sSFR and metallicity 
of $0.04<z<0.12$ SFGs with different stellar mass ranges.}
\end{center}
\end{figure*}

In Figure 2(b), we compare the SFR distributions of all samples for the 
two redshift ranges under the luminosity limit of $\rm log(L_{\OIII})>39.7$.
Following Figure 2(a), the green and black lines represent the SFR 
distributions of the SFGs at 
$0.04<z\leqslant0.06$ and $0.10<z<0.12$, respectively. The red
dotted line shows the SFR distribution of the control sample,
which has the same SFR distribution as the
$0.04<z\leqslant0.06$ SFGs, and comes randomly from the SFGs at
$0.10<z<0.12$. The blue dashed line represents the SFR distribution 
of the control sample, which has the same SFR distribution as the
$0.10<z<0.12$ SFGs, and comes randomly from the SFGs 
at $0.04<z\leqslant0.06$.

In Table 1, we show the summary of the evolution of MZ relations 
for the different samples. Under the luminosity 
threshold of $\rm log(L_{H \alpha})>41.0$ 
(also see Figure 2(a)), the median value of log(SFR), 
log($M_{\star}$), and 12+log(O/H)
is $0.36\pm0.19$, $9.40\pm0.09$, and $8.71\pm0.11$, 
respectively, in the $0.04<z\leqslant0.06$ 
SFGs with $9.2<$ log$(M_{\star}/M_{\sun})<9.5$. 
The median value of log(SFR), log($M_{\star}$), and 12+log(O/H)
is $0.37\pm0.19$, $9.41\pm0.08$, and $8.57\pm0.12$, 
respectively, in their corresponding control 
sample with $9.2<$ log$(M_{\star}/M_{\sun})<9.5$. 
The median value of 12+log(O/H) in the
$0.04<z\leqslant0.06$ SFG sample is higher than the control 
sample by $\sim 0.1$ dex. 
The Kolmogorov-Smirnov (K-S) test of the SFR and 
log($M_{\star}$) distributions 
gives rather high probabilities of $54\%$ and $83\%$, showing the
SFR and log($M_{\star}$) distributions between the two 
redshift ranges are drawn from the same population. The K-S test
of the metallicity distribution yields a very low probability of
$0.0046\%$, or equally a rejection at $99.9954\%$ confidence level,
for the null hypotheses that the two distributions are drawn from 
the same population. These imply that the MZ relation evolution is 
found in the $0.04<z\leqslant0.06$ SFG sample and its corresponding 
control sample with $9.2<$log$(M_{\star}/M_{\sun})<9.5$.
In the $0.10<z\leqslant0.12$ SFGs with 
$9.2<$log($M_{\star}/M_{\sun})<9.5$, the median value of 
log(SFR), log($M_{\star}$), and 12+log(O/H) is 
$0.42\pm0.32$, $9.42\pm0.08$, and $8.60\pm0.13$, 
respectively. The median value of log(SFR), 
log($M_{\star}$), and 12+log(O/H) is 
$0.37\pm0.17$, $9.40\pm0.09$, and $8.71\pm0.09$, 
respectively, in their corresponding control sample.
Their probabilities for the two redshift range samples are 
$53\%$, $85\%$, and $0.0296\%$, respectively. 
These indicate that the two samples with the same 
distribution in SFR and mass show significant redshift 
evolution of MZ relation.

Under the luminosity threshold of $\rm log(L_{\OIII})>39.7$ 
(also see Figure 2(b)), we follow the method of Figure 2(a). 
The median value of log(SFR), log($M_{\star}$), and 12+log(O/H) 
is $-0.11\pm0.21$, $9.35\pm0.09$, and $8.76\pm0.10$, 
respectively, in the $0.04<z\leqslant0.06$ SFGs with the same 
stellar mass range. The median value of log(SFR), 
log($M_{\star}$), and 12+log(O/H) 
is $-0.15\pm0.15$, $9.39\pm0.09$, and $8.78\pm0.11$, 
respectively, in their 
corresponding control sample. The median
value of 12+log(O/H) in the two samples has no difference 
($\sim 0.0$ dex). The K-S test of the 
SFR, log($M_{\star}$), and metallicity 
distributions yields rather high probabilities of $74\%$, $22\%$, 
and $51\%$, showing that their distributions between the two 
samples are randomly drawn from the same parent population, and 
these indicate that the $0.04<z\leqslant0.06$ sample
and its corresponding control sample do not demonstrate the
redshift evolution of the MZ relation. 
The $0.10<z<0.12$ SFGs with the same stellar mass range 
show that the median value of 
log(SFR), log($M_{\star}$), and 12+log(O/H)
is $0.27\pm0.36$, $9.40\pm0.08$, $8.66\pm0.14$, respectively. 
The median value of log(SFR), log($M_{\star}$), and
12+log(O/H) is $0.22\pm0.24$, $9.41\pm0.09$, and $8.78\pm0.11$, 
respectively, in their corresponding control sample. The
probabilities for the two redshift range samples are $65\%$, $68\%$, 
and $0.00142\%$. The K-S test
shows the redshift evolution of the MZ relation.

\begin{figure*}
\begin{center}
\includegraphics[width=14cm,height=10.5cm]{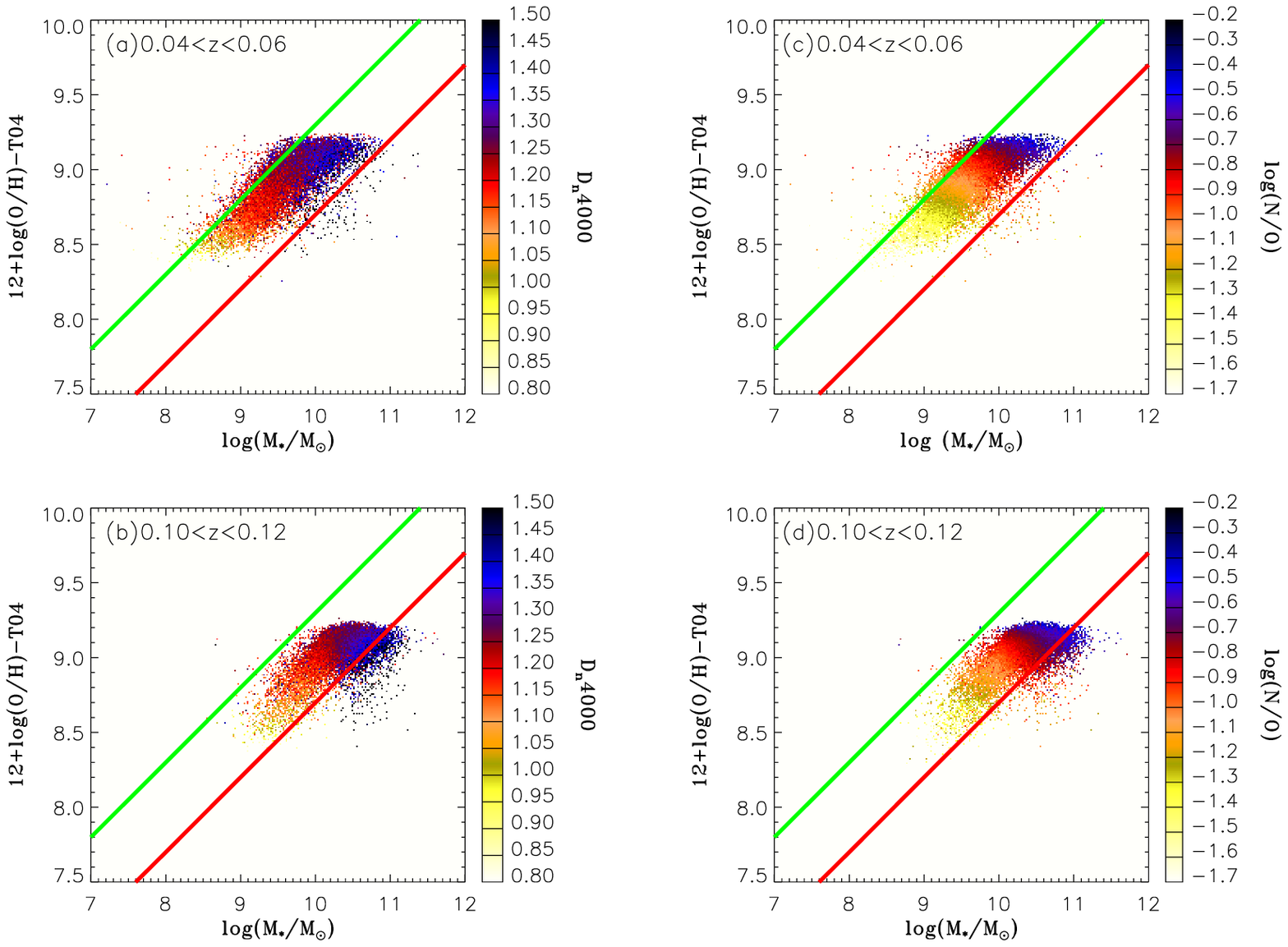}
\caption{Comparison of the MZ relations of $0.04<z\leqslant0.06$
and $0.10<z<0.12$ SFGs using the colour bars of $D_{n}4000$
and log(N/O), respectively. The green and red lines are the same diagnostic 
line as in Fig. 5.}
\end{center}
\end{figure*}

From the above results, we find that the two different samples
having higher SFRs (e.g. median(log(SFR))$>0.0$)
based on the $0.04<z\leqslant0.06$ and $0.10<z<0.12$ SFGs often 
have $\sim 0.1$ dex difference in metallicity, and that the two 
samples having lower SFRs (e.g. median(log(SFR))$<0.0$)
based on the two redshift range SFGs usually have 
no significant difference in metallicity. 
In fact, these may originate from the different SFG samples
selected different thresholds of emission line luminosities. 
In Figure 2(b), the control sample (red dotted lines) 
of the $0.04<z\leqslant0.06$ SFGs with 
$9.2<$log$(M_{\star}/M_{\sun})<9.5$ almost comes from the 
$0.10<z<0.12$ SFGs with the same stellar mass range and the smallest 
SFRs. We can see that this sample comes 
mainly from the SFGs with $\rm 12+log(O/H) >8.7$ in Figure 4(b) of Wu
et al. (2016, see the SFGs at $0.10<z<0.12$ (the red contour) and 
the luminosity threshold of $\rm log(L_{\OIII})>39.7$ (the black line)).
Although our sample is not just the same as the 
sample of Wu et al. (2016), both samples utilise the same 
luminosity threshold of $\rm log(L_{\OIII})>39.7$. 
Due to the speciality of the sample, almost no evolution of
the MZ relation is found in the SFGs with lower SFRs.
Therefore, we speculate that the evolution of the MZ relation is clearly 
observed in the higher SFR SFGs.

In addition, we study the metallicity difference 
between the two redshift range samples 
with different galaxy morphologies. 
In Figures 3(a) and 3(b), we show the metallicity distributions
of different SFGs with $9.2<$log$(M_{\star}/M_{\sun})<9.5$. 
The black line of Figure 3(a) (Figure 3(b)) is the metallicity 
distribution of $0.04<z\leqslant0.06$ ($0.10<z<0.12$) SFGs
with $9.2<$log$(M_{\star}/M_{\sun})<9.5$.
The red dotted line of Figure 3(a) (Figure 3(b)) is the metallicity
distribution of $0.04<z\leqslant0.06$ ($0.10<z<0.12$) SFGs 
with $n<1.5$ 
and $9.2<$log($M_{\star}/M_{\sun})<9.5$. From Figures 3(a)
and (b), we can see a significant difference of the metalicity 
distributions, showing that the lower redshift SFGs have higher
metallicities than the higher redshift SFGs, and that the 
median metallicity is $8.85\pm0.11$ and $8.72\pm0.13$, respectively.

In Figures 3(c) and 3(d), the black lines are the metallicity 
distributions of $0.04<z\leqslant0.06$ and $0.10<z<0.12$ SFGs 
with $9.2<$log$(M_{\star}/M_{\sun})<9.5$, respectively. 
Also, Figure 3(a) (Figure 3(b)) has the same black line as 
Figure 3(c) (Figure 3(d)). The red dotted line of Figure 3(c) 
(Figure 3(d)) is the metallicity distribution of 
$0.04<z\leqslant0.06$ ($0.10<z<0.12$) SFGs with $n>2.5$ 
and $9.2<$log$(M_{\star}/M_{\sun})<9.5$. 
From Figures 3(c) and 3(d), a clear difference of the metalicity 
distributions is presented, showing that the lower redshift SFGs have 
higher metallicities than the higher redshift SFGs, and that the 
median metallicity is $8.78\pm0.12$ and $8.60\pm0.14$, 
respectively. This
indicates that the evolution of the MZ relation always holds in
elliptical or spiral galaxies.

In Figures 4(a) and 4(b), the black lines are the metallicity 
distributions of $0.04<z\leqslant0.06$ and 
$0.10<z<0.12$ SFGs with $9.2<$log$(M_{\star}/M_{\sun})<9.5$, 
respectively. 
The red dotted line of Figure 4(a) (Figure 4(b)) shows the metallicity 
distribution of $0.04<z\leqslant0.06$ ($0.10<z<0.12$) 
SFGs with $c<2.86$ and $9.2<$log$(M_{\star}/M_{\sun})<9.5$. From 
Figures 4(a) and 4(b), we can see a significant difference 
between the metalicity distributions, showing that the lower redshift 
SFGs have higher metallicities than the higher redshift SFGs, and that 
the median metallicity is $8.83\pm0.12$ and $8.67\pm0.14$, 
respectively.

In Figures 4(c) and 4(d), the black lines are the 
metallicity distributions of $0.04<z\leqslant0.06$ and 
$0.10<z<0.12$ SFGs with 
$9.2<$log$(M_{\star}/M_{\sun})<9.5$, respectively. 
Moreover, Figure 4(a) (Figure 4(b)) has the same black line as Figure 4(c) 
(Figure 4(d)). The red dotted line of Figure 4(c) (Figure 4(d)) presents the 
metallicity distribution of $0.04<z\leqslant0.06$ ($0.10<z<0.12$) SFGs
with $c\geqslant2.86$ and $\rm 9.2<log(\rm M_{\star}/M_{\sun})<9.5$. 
A significant difference between the metallicity distributions is shown 
in Figures 4(c) and 4(d), presenting that the lower redshift SFGs have 
higher metallicities than the higher redshift SFGs, and that the
 median metallicity is $8.77\pm0.13$ and $8.60\pm0.14$, 
respectively. These show that the MZ relation evolution should not 
depend on the elliptical or spiral galaxies.

\begin{figure*}
\begin{center}
\includegraphics[width=16cm,height=12cm]{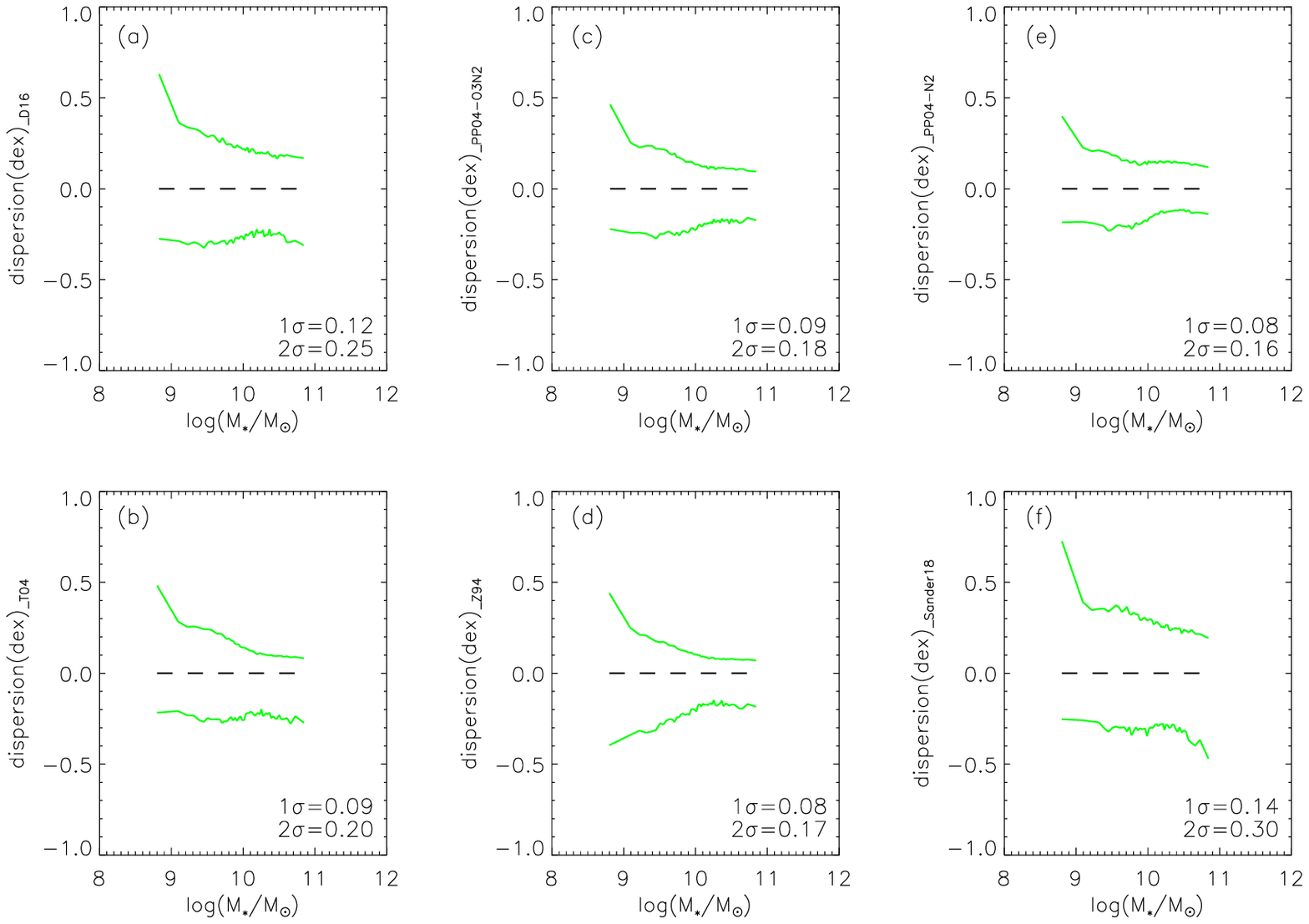}
\caption{The scatter of the MZ relations of $0.04<z<0.12$ SFGs 
for the different metallicity calibrators. The green
solid curves are the $95\%$ contours of metallicity dispersion
in all MZ relations.}
\end{center}
\end{figure*}

\subsection{The dependence of the MZ relation}

In this section, we first utilise the concentration 
index to investigate
the morphology dependence of the MZ relation. Then, we study the
morphological classifications of the MZ relation using the
S\'{e}rsic index to separate galaxies into elliptical and
spiral galaxies. Moreover,  we explore the SFR and sSFR dependence of 
the MZ relation. Finally, we further investigate the 
dependence on the stellar population age of the MZ relation 
using the $D_{n}4000$ index and log(N/O).

In Figures 5(a) and 5(b), we show the MZ relations of
$0.04<z\leqslant0.06$ and $0.10<z<0.12$ SFGs with colour bars of
concentration index, respectively. The colour bars show 
the c values per
bin (0.1 dex in log($M_{\star}/M_{\sun}$) and 0.05 dex in
metallicity) in the MZ relation. The green and red lines 
are the diagnostic lines
for comparing MZ relations between the higher and lower redshift
SFGs. Compared to the MZ relation of Figure 5(b), the scatter at 
a lower galaxy stellar mass is exhibited in Figure 5(a), showing the trend
of decreasing scatter with increasing stellar mass 
(Tremonti et al. 2004; Zahid et al. 2012). The origin of 
the scatter may originate from
accretion and/or mergers (Bothwell et al. 2013; Forbes et al. 2014),
and Kacprazk et al. (2016) suggested that an accretion is likely to
have a great influence on a galaxy's mass.

Compared to the $0.10<z<0.12$ SFGs, the $0.04<z\leqslant0.06$ SFGs
have higher metallicities at a fixed stellar mass. At
a fixed stellar mass, there is a tendency that the concentration
index decreases as the metallicity increases, while at a fixed 
metallicity, the concentration index increases as the galaxy 
stellar mass increases. The index is often used to categorize
galaxies into early-type and late-type with the different recommended
values (Shimasaku et al. 2001; Nakamura et al. 2003; Shen et al.
2003). Hooper et al. (2007) and Ellison 
et al. (2008) respectively reported the dependence of 
the MZ relation on galaxy
size, which is related to the morphology structure. Therefore the 
MZ relation should depend on the galaxy structure properties.

In Figures 5(c) and 5(d), 
the MZ relations of $0.04<z\leqslant0.06$
and $0.10<z<0.12$ SFGs are shown using colour bars of
S\'{e}rsic index, respectively. The green and red lines are
the same diagnostic lines as in Figures 5(a) and 5(b). At a fixed 
stellar mass, there is a trend that the S\'{e}rsic index 
decreases with increasing metallicity, while at a fixed metallicity, 
the index increases with increasing stellar mass, and the change of 
different structural types is from 
disk-dominated galaxeies ($n<1.5$) to bulge-dominated galaxies 
($n>2.5$, Maier et al. 2009). From Figures 5(a) and 5(b), we can see a
significant trend that the concentration index almost 
does not change when
the stellar mass and metallicity simultaneously increase. The same 
trend for S\'{e}rsic index also exists in Figures 5(c) and 5(d), 
and it also appears at the log$(M_{\star}/M_{\sun})<10.0$ 
in Figures 5(e) and 5(f) (with regard to the MZ relation of 
log$(M_{\star}/M_{\sun})>10.0$ SFGs, 
we will discuss it in next paragraph.), suggesting that the trend of 
Figures 5(e) and 5(f) is a result of some dependence of 
the MZ relation on SFR. 
In Figure 5, when a concentration index or Sersic index of the MZ 
relation increases, its SFR also increases. This indicates that 
galaxy structure properties are closely related to SFRs. 
In addition, we can see another significant tendency that the SFRs increase 
with increasing $M_{\star}$ and Z simultaneously at the 
log$(M_{\star}/M_{\sun})>10.0$ in Figures 5(e) and 5(f), in other 
word, high stellar
mass galaxies with higher SFRs often have higher metallicities. This 
result is similar to one of S\'{a}nchez et al (2017), who studied 
the MZ relation with 734 CALIFA galaxies.

\begin{figure*}
\begin{center}
\includegraphics[width=14cm,height=10.5cm]{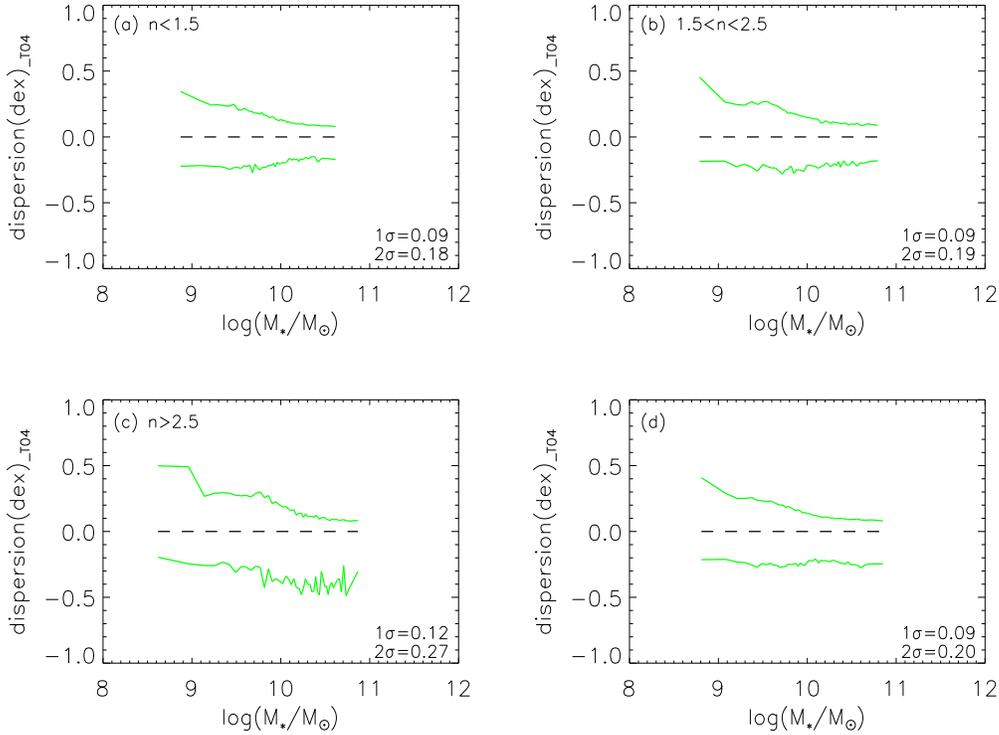}
\caption{Comparison of the scatter of MZ relations 
of $0.04<z<0.12$ SFGs for the different S\'{e}rsic indexes. 
The symbols are the same as in Fig. 9.}
\end{center}
\end{figure*}

In Figures 5(e) and 5(f), the log$(M_{\star}/M_{\sun}) < 10.0$ and 
log$(M_{\star}/M_{\sun}) > 10.0$ SFGs present a different 
relation between SFR and Z at a fixed stellar mass; the former population
shows that the metallicity increases with decreasing SFR, whereas the latter
population presents that the metallicity increases with increasing SFR. We 
suggest that the two group SFGs may come from differen galaxy populations, 
and that they are likely to possess different stellar population ages.

In Figures 5(e) and 5(f), 
the two galaxy populations seem to be observed in 
log$(M_{\star}/M_{\sun}) \sim 10.0$, so we show the MZ relations using 
the six metallicity estimators (see the first paragraph of Section 3.3) 
and colour bars of sSFR in Figures 6(a) -- 6(f), presenting a 
continuous change in sSFR in log$(M_{\star}/M_{\sun}) >10.0$ SFGs. 
The sample shown in Figure 6 includes all SFGs with $0.04<z<0.12$. 
Whether do the two galaxy populations of Figure 5 exist in Figure 6, so we 
use the sample of 81,125 SFGs to present the MZ relation in each plot.
Figure 6 displays an anticorrelation, as a whole, 
that galaxies with higher sSFRs have lower metallicities at a fixed stellar 
mass. Actually, we seem to see the two galaxy populations; one 
presents an anticorrelation between Z and sSFR in the lower mass galaxies, and 
another one exhibits a weak positive correlation between Z
and sSFR in the higher mass galaxies, showing a large scatter, and this 
is consistent with 
Lara-L\'{o}pez et al. (2013). But Salim et al. (2014) suggested that the 
metallicity calibrator may lead to the positive correlation between 
metallicity and sSFR, 
and that the latter galaxy population is not exist. Actually, we show 
the second galaxy population (see the galaxies at log(sSFR)$\lesssim-10.2$) 
in Figures 6(a)--6(f), showing independence on the metallicity estimator. 
Due to the larger stellar masses and smaller sSFRs from these figures, 
we suggest that these galaxies may have older stellar  populations (see Figures 
8(a) and 8(b)).

To show clearly the two galaxy populations, Figure 7
displays the graphs of Z versus sSFR in SFGs with different 
stellar mass ranges. In Figures 7(a) and 7(b), we demonstrate 
the relations between sSFR and 12+log(O/H)
in SFGs with log$(M_{\star}/M_{\sun})<10.0$ and 
log$(M_{\star}/M_{\sun})>10.0$, respectively.
From Figure 7(a), an anticorrelation is significant 
(Spearman's coefficient, r=-0.62). 
In Figure 7(b), since the SFG sample with 
log$(M_{\star}/M_{\sun})>10.0$
are mixed with some SFGs having an anticorrelation 
between sSFR and 12+log(O/H),
 their correlation is not clear, showing their Spearman coefficient r=0.14.
Figures 7(c) and 7(d) display the relations between sSFR and 12+log(O/H) 
in SFGs with
log$(M_{\star}/M_{\sun})<10.5$ and log$(M_{\star}/M_{\sun})>10.5$, 
respectively. Compared to Figure 7(a), $10.0<$log$(M_{\star}/M_{\sun})<10.5$
SFGs are mixed into the SFG sample 
with log$(M_{\star}/M_{\sun})<10.5$ in Figure 7(c), and
their correlation has weakened, showing their Spearman coefficient r=-0.47. 
Compared to Figure 7(b), the correlation
significantly increases in Figure 7(d), showing  their Spearman coefficient 
r=0.44. The two correlations are supported by some recent
findings (Yates, Kauffmannn, \& Guo 2012; Andrew \& Martini 2013; 
Yates \& Kauffmann 2014). Regarding the positive correlation at high 
mass galaxies, it should be investigated by more observational studies.

\begin{figure*}
\begin{center}
\includegraphics[width=8cm,height=6cm]{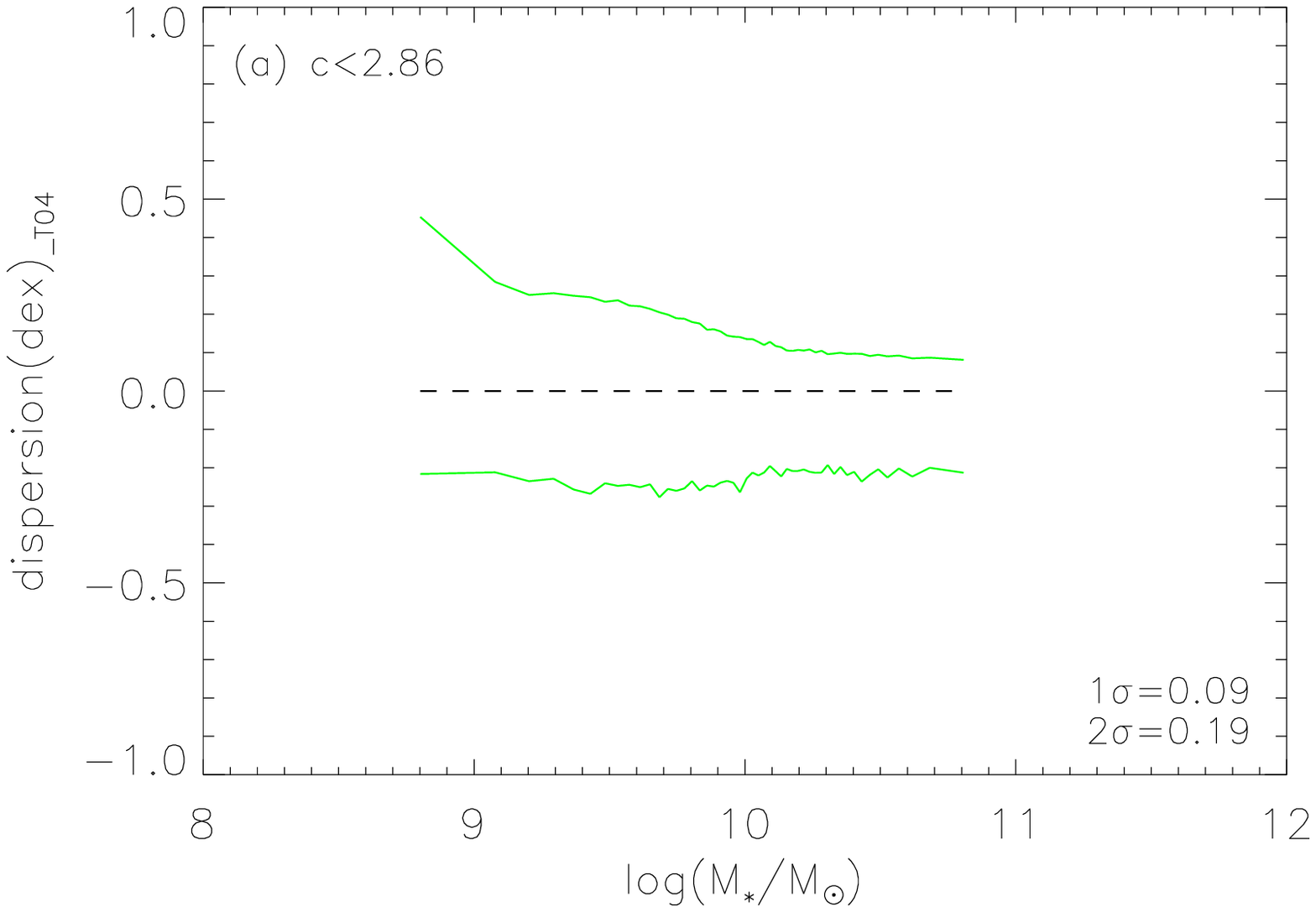}
\includegraphics[width=8cm,height=6cm]{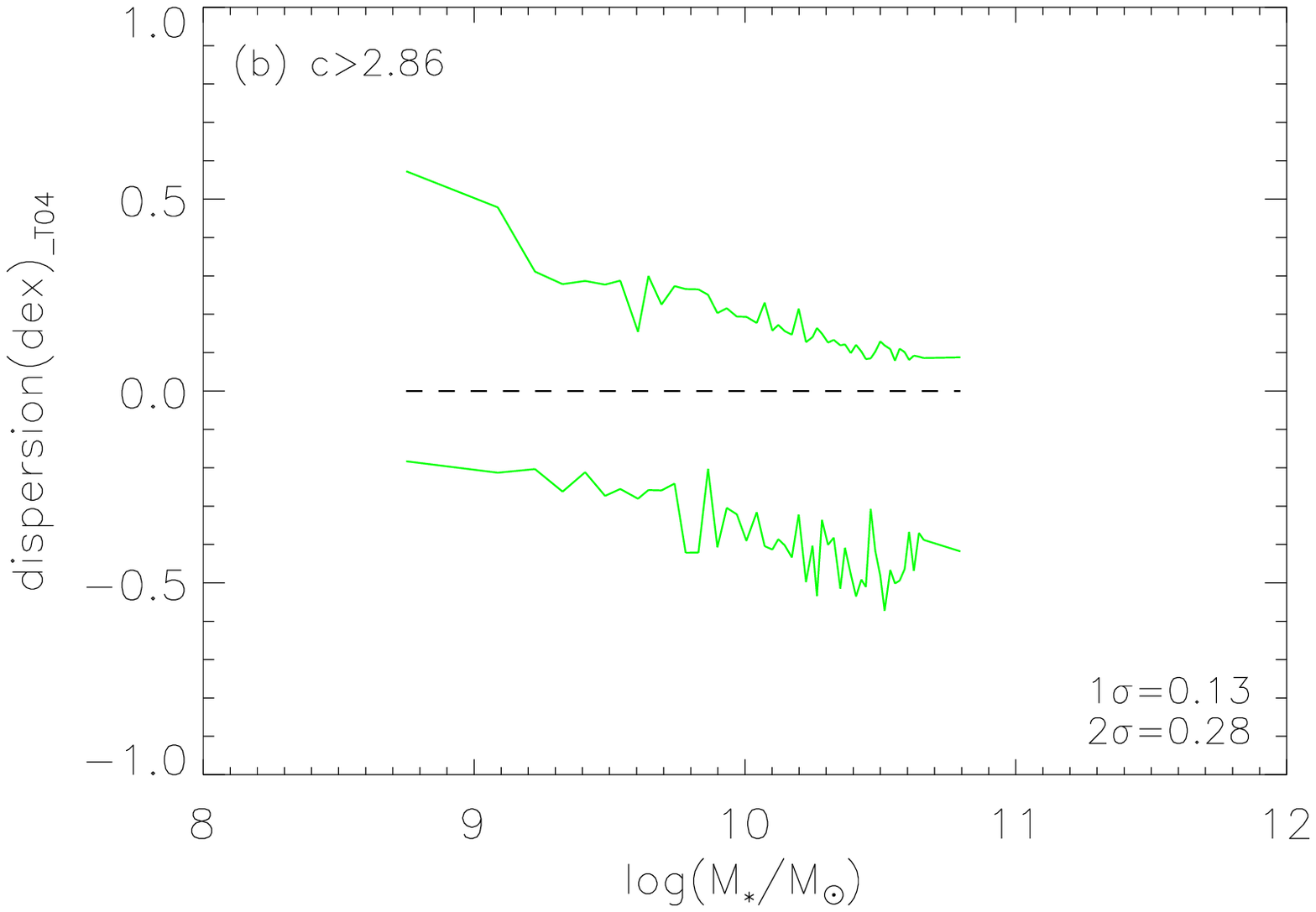}
\caption{Comparison of the scatter of MZ relations of $0.04<z<0.12$ SFGs for the
different concentration indexes. The symbols are the same as  
in Fig. 9.}
\end{center}
\end{figure*}

It is well known that a $D_{n}4000$ index is a good proxy for the
mean stellar ages of stellar population below $\sim2$ Gyr. In Figures
8(a) and 8(b), we show the MZ relations of $0.04<z\leqslant0.06$ and
$0.10<z<0.12$ SFGs using the colour bars of $D_{n}4000$ index,
respectively. The green and red lines are the same diagnostic lines 
as in Figure 5. As can be seen from Figures 8(a) and 8(b), galaxies with 
higher $D_{n}4000$ usually have larger S\'{e}rsic
 indexes at a given stellar mass (Zahid \& Geller 2017).

From Figure 8(a), we can see a trend that the $D_{n}4000$ index
increases when the stellar mass and metallicity increase, and this is
a significant difference compared to the MZ relations with the colour
bars of concentration index and S\'{e}rsic
index in Figure 5, indicating that the MZ relation may be related
to the stellar age of SFGs. Using a sample of Lyman-break 
analogure, Lian et al. (2015) found that 
the MZ relation is strongly correlated with $\rm D_{n}4000$, showing
that galaxies having higher $D_{n}4000$ present higher 
metallicities at a given stellar mass. Using the similar method of 
Kauffmann et al. (2003b), Wu \& Zhang (2013) 
presented a nice relation between 
log($M_{\star}$) and log(N/O), and suggested 
that the ratio may be used as a standard 
candle (Wu \& Zhang 2013). In Figures 8(c) and 8(d),  we find a clear 
tendency that the log(N/O) increases with increasing $M_{\star}$ and Z, 
implying that the MZ relation should closely link the stellar age. 
Actually, the tendency should be understandable.
Since this $D_{n}4000$ index can provide a diagnostic of 
the past star formation histories of galaxies 
(Kauffmann et al. 2003a), we obtain the age imformation 
of different processes during galaxy evolution. With galaxy 
evolution, a huge amounts of existed or inflow gas is 
transformed into stars, which results in increasing galaxy stellar 
mass. At the same time, the evolution of massive 
stars and intermediate-mass stars in galaxies contributes 
increasing of 12+log(O/H) and log(N/O) ratio (Wu et al. 2013).
As can be seen, this relationship between the stellar mass and 
metallcity almost accompanies the change of $D_{n}4000$ and 
log(N/O) ratio during the whole galaxy evolution. We confirm 
that the MZ relation may link closely the 
stellar age in SFGs (Gallazzi et al. 2005; Lian et al. 2015).

Combining with Figures 5(a), 5(b), 5(c), and 5(d), we find that the 
correlations between $M_{\star}$ and Z are observed 
at a certain range of the concentration indexes or S\'{e}rsic indexes. 
From Figure 5, we can see the metallicity tends to be lower in 
galaxies with a higher concentration index, higher S\'{e}rsic index, or
higher SFR. In Figures 5(e) and 5(f), we seem to see the 
two galaxy populations at 
log$(M_{\star}/M_{\sun})\sim 10.0$, and we find that the two populations 
are observed by using the colour bars of sSFR in Figures 6(a) -- 6(f). 
In addtion, Figure 7 displays the two galaxy populations, presenting 
clearly the two correlations between sSFR and Z, and 
the latter galaxy population does not depend on the metallicity estimator.
We can see that the stellar mass and metallicity usually present
higher in galaxies with a higher $D_{n}4000$ or higher log(N/O) ratio from
Figure 8.

\begin{figure*}
\begin{center}
\includegraphics[width=16cm,height=12cm]{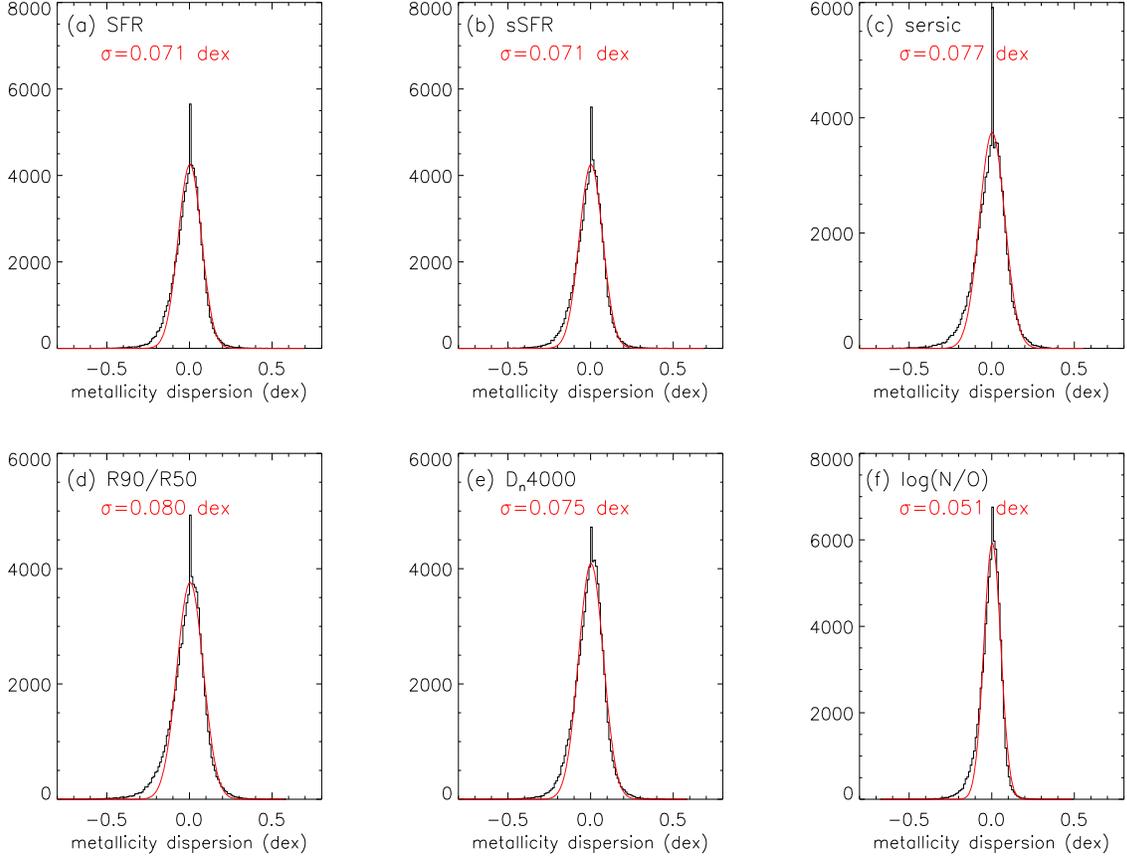}
\caption{Comparison of metallicity dispersion in MZ-X relation, 
X is the third parameter (SFR, sSFR, sersic index, concentration 
index, $D_{n}4000$, and log(N/O), respectively.). The histogram displays
the difference from the median calculated in bins of 0.05 dex in $M_{\star}$
and X. The red curve is a Gaussian with $\sigma$.}
\end{center}
\end{figure*}

\subsection{The scatter of the MZ relation}

In Figure 9, we show  the scatter of the MZ relation of more than
$80,000$ galaxies using the six metallicity-calibrated 
methods, such as 
the metallicity diagnostics of Z94, PP04-O3N2, PP04-N2, T04, Z94, and 
Sander18. Firstly, the sample is classifyed into 50 almost equal 
galaxy bins in stellar mass, then we obtain the median value of 
the metallicities (get two subbins based on the median metallicity) 
and stellar masses, and calculate respectively the $95\%$ contours of
metallicity dispersion (deviation from the median metallicity) of both 
subbins in each bin. The contours are shown by the green solid lines 
in Figure 9.

It is observed that the scatter increases with decreasing galaxy 
stellar mass in Figure 9 (Zahid et al. 2012). Whether the trend is real? 
Zahid et al. (2012) checked it using the five diagnostic 
calibrators of metallicities in their Figure 5, and we also 
examine it using another
five methods of metallicity estimators in Figure 9. We find that the
trend does not depend on the metallicity calibrator. Moreover, we 
find that the smallest scatter ($2\sigma$) is 0.16 dex in 
log[([N~II] $\rm \lambda 6584)]/H\alpha]$ (N2) of PP04, and  the 
largest one ($2 \sigma$) is 0.30 dex in 
log[([N~II]$\lambda 6584)$]/([O~II] $\lambda 3726, 3729)]$ (N2O2) of 
Sanders et al. (2018).

In Figures 9(a), 9(c), and 9(e), the scatter is almost symmetric for galaxies 
with $\sim 9.5 <$ log$(M_{\star}/M_{\sun})<10.5$, while it is clearly 
asymmetric in Figures 9(b), 9(d), and 9(f). In Figure 4 of Zahid et al. (2012), 
the scatter shown is symmetric at 
$8.5<$log$(M_{\star}/M_{\sun})<9.5$ galaxies, 
and we find that the scatter symmetry may occur at the different stellar 
masses, determined by the different metallicity 
calibrator. At log$(M_{\star}/M_{\sun})>10.5$, 
almost all the scatter 
in Figures 9 begin to be asymmetric. Compared to the 
lower dispersion below the median values, the upper dispersion 
above the median values
decreases significantly with the galaxy stellar mass in Figures 
9(a), and 9(f), and the dispersion seems to do not decrease in Figures 
9(b) and 9(d). In addition, Figures 9(a) and 9(f) show larger 
upper dispersion at log$(M_{\star}/M_{\sun})\sim 11.0$, showing
no saturation in metallicity in the Dopita (2016) and Sanders et al. (2018)
diagnostic calibrators.

In Figure 10, we display the scatter of MZ relations of SFGs with the 
different S\'{e}rsic indexes. Figures 10(a) and 10(b) show the 
scatter of MZ relations of SFGs with $n<1.5$ and $1.5<n<2.5$, 
respectively. Their sample sizes are 41,059 and 28,310,
presenting $2\sigma=0.18$ dex and $2\sigma=0.19$ dex, respectively. 
Figures 10(c) and 10(d) present the scatter of MZ relations of 
SFGs with $n>2.5$ and the total sample, respectively. Their sample 
sizes are 11,564 and 80,933 (192 galaxies at $n=0$), 
presenting $2\sigma$ scatter of $0.27$ 
dex and $0.20$ dex, respectively. From Figure 10, we find 
that the scatter of the MZ relation colsely links the galaxy 
morphology, showing that elliptical galaxies have larger 
scatter than spiral galaxies. 
Certainly, the large scatter in the sample with $n>2.5$ 
in which the observational uncertainties in the metallicity 
measurement may be much higher than other two samples.

In Figure 11, the scatter of MZ relations of SFGs with the 
different concentration indexes is shown. Figures 11(a) and 11(b)
display the scatter of MZ relations of SFGs with 
$c<2.86$ and $c\geqslant2.86$, 
respectively. Their sample sizes are 76,286 and 4,773 (66 galaxies at c=0), 
presenting  $2\sigma=0.19$ dex and 
$2\sigma=0.28$ dex, 
respectively. Figure 11 confirms the result 
of Figure 10, and the scatter of the MZ relation is closely related the 
galaxy morphology. In addition, we show that the scatter 
is composed of the superposition of several scatter of MZ relations
for different type galaxies of the sample.

In addition, we utilise the scatter to compare the dependence 
of metallicity on galaxy properties other than mass. Figure 9(b) 
displays that the scatter in MZ relation is $\sigma=0.09$ dex.
In Figure 12, we show metallicity dispersion in MZ - X relation, 
and X is the third parameter. The dispersion is calculated in bins 
of 0.05 dex in $M_{\star}$ and X. The dispersion reduces to 
$\sim 0.07$ dex when adopting this SFR paremeter in the MZ relation
(see Figure 12(a)). In Figure 12(b), 12(c), 12(d), 12(e), and
12(f), the scatter decreases to $\sim0.07$ dex, $\sim0.08$ dex, 
$\sim0.08$ dex, $\sim0.07$ dex, and $\sim0.05$ dex, respectively,
 when sSFR, Sersic index, concentration index, $D_{n}4000$, 
and log(N/O) are used as the third parameter. We can see a
decease in scatter when adopting the third parameter in MZ relations.
Compared to the scatter of $\sigma \sim 0.05$ in Mannuccci et al. (2010),
the dispersion in MZ-SFR relation is $\sigma \sim 0.07$, and this 
difference
may be related to the different SFG sample. In addtion, we find 
that the dispersion in MZ-log(N/O) relation is $\sigma \sim 0.05$, 
showing the smallest dispersion among these galaxy properties, and
this may originate from that the log(N/O) ratio is used to calculate 
the SFG metallicity.

\section{Summary}

Utilising the observational data of 86,111 SFGs obtained from the
catalogue of the SDSS DR7 MPA-JHU emission-line measurements, we
investigate the MZ relation evolution. Also, using the
concentration index, S\'{e}rsic index, selected both 
from the NYU-VAGC data, log(N/O), SFR, sSFR, and $D_{n}4000$,
we explore the dependence and scatter 
of the MZ relations of $0.04<z\leqslant0.06$ and $0.10<z<0.12$ SFGs. 
Our main results are summarized as followings:

1. Under the $\rm log (L_{H \alpha})>41.0$ threshold, the
$0.04<z\leqslant0.06$ ($0.10<z<0.12$) SFGs with 
$9.2<\rm log(M_{\star}/M_{\sun})<9.5$ have higher (lower) $\sim 0.1$ 
dex in metallicity than their corresponding control
sample, and the K-S test shows that the two samples 
(both median(log(SFR))$>0.0$) present the MZ relation evolution.

2. Under the $\rm log(L_{\OIII})>39.7$ threshold, the
$0.04<z\leqslant0.06$ SFGs with 
$9.2<\rm log(M_{\star}/M_{\sun})<9.5$ have
almost no difference in metallicity compared to 
their corresponding control
sample, and the K-S test ($p_{null}=51\%$, see Table 1)
shows that the MZ relation evolution is not found in the 
two samples (both median(log(SFR))$<0.0$). However, the $0.10<z<0.12$ 
SFGs with $9.2<\rm log(M_{\star}/M_{\sun})<9.5$ have the same result 
as the $\rm log(L_{H \alpha})>41.0$ threshold mentioned 
above ($p_{null}=1.42\times10^{-3}\%$, see Table 1), 
compared to their corresponding control sample 
(both median(log(SFR))$>0.0$). These show that the evolution of 
the MZ relation is clearly observed in the higher SFR SFGs 
(e.g. median(log(SFR))$>0.0$)), and that almost no evolution of the 
MZ relation is found in the SFGs at lower SFRs (e.g. median(log(SFR))$<0.0$).
We suggest that these originate from the speciality of the sample.

3. Under the same stellar mass and S\'{e}rsic index 
(concentration index), we
show that the $0.04<z\leqslant0.06$ SFGs have always higher 
$\sim 0.1 - 0.2$ dex in metallicity than the $0.10<z<0.12$ SFGs.
This shows that the MZ relation evolution exists and does not depend on the 
elliptical or spiral galaxies.

4. Utilising the concentration index and S\'{e}rsic index,
SFR, we find that the metallicity tends to be lower in galaxies
with a higher concentration, higher S\'{e}rsic index, or higher SFR.
In addition, we can see that the stellar mass and metallicity usually present
higher in galaxies with a higher $D_{n}4000$ or higher log(N/O) ratio.

5. With regard to a possibility of the two galaxy populations in the
MZ relation at log$(M_{\star}/M_{\sun})\sim 10.0$, we find that they 
may exist by using color bars of sSFR in Figures 6(a) -- 6(f). In 
addition, we present clearly an anticorrelation and a positive 
correlation between sSFR and 12+log(O/H) at 
log$(M_{\star}/M_{\sun}) \sim 10.0$ and 10.5, respectively. 
From the dependence on the $D_{n}4000$ index and log(N/O) in the 
MZ relation, we confirm that the MZ relation may link closely 
the stellar age in SFGs.

6. The scatter of the MZ relation increases with decreasing stellar 
mass, and we show that the trend does not depend on the metallicity 
calibrator. Also, we find that the scatter symmetry may occur at 
different stellar masses, determined by the different metallicity
calibrator. Moreover, almost all the scatter of the MZ relations 
begin to be asymmetric at log$(M_{\star}/M_{\sun})>10.5$.

7. We find that the scatter of the MZ relation colsely links the galaxy
morphology, presenting that the elliptical galaxies often 
have larger scatter than the spiral galaxies. This may be originated from the 
large scatter in the SFG sample with $n>2.5$ or $c>2.86$. In addition, 
we show that the scatter is composed of the superposition of 
several scatter of the MZ relation for different type galaxies.

8. We show that metallicity dispersion in MZ - X relation, and X 
is SFR, sSFR, S\'{e}rsic index, concentration index, $D_{n}4000$,
and log(N/O), respectively. We confirm that the MZ relation 
depends on the six galaxy properties.

\section*{Acknowledgement}

We especially thank the anonymous referee for valuable
suggestions and comments, which greatly improved the paper. 
WZ acknowledges partial funding support by the Joint
Research Fund in Astronomy (U1531118) under cooperative agreement
between the NSFC and CAS.

\bsp
\label{lastpage}


\begin{thebibliography}{}



\bibitem[\protect\citeauthoryear{Abazajian}{2009}]{a} Abazajian, K. N., Adelman-McCarthy, J. K., Ag\"{u}eros, M. A. et al. 2009, \apjs, 182, 543
\bibitem[\protect\citeauthoryear{Andrews}{2013}]{Ba} Andrews, B. H., \& Martini, P. 2013, \apj, 765, 140
\bibitem[\protect\citeauthoryear{Baldwin}{1981}]{Ba} Baldwin, J. A., Phillips, M. M., \& Terlevich, R. 1981, PASP, 93, 5
\bibitem[\protect\citeauthoryear{Balogh et al}{1999}]{Ba} Balogh, M. L., Morris, S. L., Yee, H. K. C., Carlberg, R. G., \& Ellingson, E. 1999, \apj, 527, 54
\bibitem[\protect\citeauthoryear{Barrera-Ballesteros}{2017}]{Ba4} Barrera-Ballesteros, J. K., S\'{a}nchez, S. F., Heckman, T., et al. 2017, \apj, 844, 80
\bibitem[\protect\citeauthoryear{Belli}{2013}]{Ba} Belli, S., Jones, T., Ellis, R. S., \& Richard, J. 2013, \apj, 772, 141
\bibitem[\protect\citeauthoryear{Blanton}{2005}]{Bl4} Blanton, M. R. et al., 2005, AJ, 129, 2562
\bibitem[\protect\citeauthoryear{Bothwell}{2013}]{Bl4} Bothwell, M. S., Maiolino, R., Kennicutt, R., et al. 2013, \mnras, 433, 1425
\bibitem[\protect\citeauthoryear{Brown et al. 2017}{2017}]{B2} Brown, T., Cortese, L., Catinella, B., \& Kilborn, V. 2018, \mnras, 473, 1868
\bibitem[\protect\citeauthoryear{Chabrier 2003}{2003}]{Ch0} Chabrier, G. 2003, PASP, 115, 763
\bibitem[\protect\citeauthoryear{Christensen 2012}{2012}]{Ch2} Christensen, L., Richard, J., Hjorth, J., et al. 2012, \mnras, 427, 1953
\bibitem[\protect\citeauthoryear{Cullen et al. 2014}{2014}]{C1} Cullen, C., Cirasuolo, M., McLure, R. J., Dunlop, J. S., \& Bowler, R. A. A. 2014, \mnras, 440, 2300
\bibitem[\protect\citeauthoryear{Dopita 2016}{2016}]{D3} Dopita, M, A. 2016, APSS, 361, 61
\bibitem[\protect\citeauthoryear{Ellison et al. 2008}{2008}]{e0} Ellison, S. L., Patton, D. R., Simard, L., \& McConnachie, A. W. 2008, \apjl, 672, 107
\bibitem[\protect\citeauthoryear{Erb et al. 2006}{2006}]{e1} Erb, D. K., Shapley, A. E., Pettini, M., et al. 2006, \apj, 644, 813
\bibitem[\protect\citeauthoryear{Forbes et al. 2014}{2014}]{F0} Forbes, J. C., Krumholz, M. R., Burkert, A., \& Dekel, A. 2014, \mnras, 443, 168
\bibitem[\protect\citeauthoryear{Gallazzi et al. 2005}{2005}]{G0} Gallazzi, A., St\'{e}phane, C., Brinchmann, J., et al. 2005, \mnras, 362, 41
\bibitem[\protect\citeauthoryear{Guo et al. 2016}{2016}]{G1} Guo,Y.C., et al. 2016, \apj, 822, 103
\bibitem[\protect\citeauthoryear{Henry et al. 2013}{2013}]{He1} Henry, A., Scarlata, C., Dom\'{\i}nguez, A., Malkan, M., Martin, C. L., et al. 2013, \apjl, 776, 27
\bibitem[\protect\citeauthoryear{Hoopes 2007}{2007}]{Hoo1} Hoopes, C. G., Heckman, T. M., Salim, S., et al. 2007, \apjs, 173, 441
\bibitem[\protect\citeauthoryear{Hunt 2016}{2016}]{Hu1} Hunt, L., Dayal, P., Magrini, L., \& Ferrara, A. 2016, \mnras, 463, 2002
\bibitem[\protect\citeauthoryear{Juneau 2014}{2014}]{J1} Juneau S. et al., 2014, \apj, 788, 88
\bibitem[\protect\citeauthoryear{Kacprzak et al. 2016}{2016}]{ka0} Kacprzak, G. G., van de Voort, F., Glazebrook, K., et al. 2016, \apjl, 826, L11
\bibitem[\protect\citeauthoryear{Kashino et al. 2017}{2017}]{Ka} Kashino, D., Silverman, J. D., Sanders, D., et al. 2017, \apj, 835, 88
\bibitem[\protect\citeauthoryear{Kauffmann et al. 2003}{2003}]{Ka1} Kauffmann, G., Heckman, T. M., Tremonti, C., et al. 2003a, \mnras, 346, 1055
\bibitem[\protect\citeauthoryear{Kauffmann et al. 2003}{2003}]{Ka2} Kauffmann, G., Heckman, T. M., White, S. D. M., et al. 2003b, \mnras, 341, 54
\bibitem[\protect\citeauthoryear{Kewley \& Dopita 2002}{2002}]{Ke0} Kewley, L. J., \& Dopita, M. A. 2002, \apjs, 142, 35
\bibitem[\protect\citeauthoryear{Kewley \& Ellison 2008}{2008}]{Ke1} Kewley, L. J., \& Ellison, S. L. 2008, \apj, 681, 1183
\bibitem[\protect\citeauthoryear{Kewley et al. 2006}{2006}]{Ke2} Kewley, L. J., Grovers, B., Kauffmann, G., \& Heckman, T. 2006, \mnras, 372, 961
\bibitem[\protect\citeauthoryear{Kewley et al. 2005}{2005}]{Ke3} Kewley, L. J., Jansen, R. A., \& Geller, M. J. 2005, PASP, 117, 227
\bibitem[\protect\citeauthoryear{Kroupa 2001}{2001}]{Kr0} Kroupa, P. 2001, \mnras, 322, 231
\bibitem[\protect\citeauthoryear{Lara-L\'{o}pez et al. 2010}{2010}]{La0} Lara-L\'{o}pez, M. A., Bongiovanni, A., et al. 2010, \aap, 519, 31
\bibitem[\protect\citeauthoryear{Lara-L\'{o}pez et al. 2013}{2013}]{La1} Lara-L\'{o}pez, M. A., Hopkins, A. M., L\'{o}pez-S\'{a}nchez, A. R., et al. 2013, \mnras, 434, 451
\bibitem[\protect\citeauthoryear{Lequeux et al. 1979}{1979}]{Le1} Lequeux, J., Peimbert,M., Rayo, J. F., Serrano, A., \& Torres-Peimbert, S. 1979, \aap, 80, 155
\bibitem[\protect\citeauthoryear{Lian et al. 2015}{2015}]{Li1} Lian, J. H., Li, J. R., Yan, W., \& Kong, X. 2015, \mnras, 446, 1449
\bibitem[\protect\citeauthoryear{Maier et al. 2009}{2009}]{Ma0} Maier, C., Lilly, S. J., Zamorani, G., et al. 2009, \apj, 694, 1099
\bibitem[\protect\citeauthoryear{Maier et al. 2014}{2014}]{Ma10} Maier, C., Lilly, S. J., Ziegler, B. L., et al. 2014, \apj, 792, 3
\bibitem[\protect\citeauthoryear{Maiolino et al. 2008}{2008}]{Ma2} Maiolino, R., Nagao, T., Grazian, A., et al. 2008, \aap, 488, 463
\bibitem[\protect\citeauthoryear{Mannucci et al. 2010}{2010}]{Ma5} Mannucci, F., Cresci, G., Maiolino, R., Marconi, A., \& Gnerucci, A. 2010, \mnras, 408, 2115
\bibitem[\protect\citeauthoryear{Nakamura et al. 2003}{2003}]{Mc0} Nakamura, O., Fukugita, M., Yasuda, N., et al. 2003, AJ, 125, 1682
\bibitem[\protect\citeauthoryear{Peng et al. 2015}{2015}]{Pe1} Peng, Y., Maiolino, R., \& Cochrane, R. 2015, Nature, 521, 192
\bibitem[\protect\citeauthoryear{Pettini \& Pagel. 2004}{2004}]{Pe1} Pettini, M., \& Pagel,B. E. J. 2004, \mnras, 348, 59
\bibitem[\protect\citeauthoryear{Pilyugin et al. 2006}{2006}]{Pi1} Pilyugin, L. S., Thuan, T. X., \& V\'{\i}lchez, J. M. 2006, \mnras, 367, 1139
\bibitem[\protect\citeauthoryear{Pilyugin et al. 2010}{2010}]{Pi2} Pilyugin, L. S., V\'{\i}lchez, J. M., Cedr\'{e}s, B., \& Thuan, T. X. 2010, \mnras, 403, 896
\bibitem[\protect\citeauthoryear{Salim et al. 2015}{2015}]{Sa0} Salim, S., Lee, J. C., Dav\'{e}, R., \& Dickinson, M. 2015, ApJ, 808, 25
\bibitem[\protect\citeauthoryear{Salim et al. 2014}{2014}]{Sa1} Salim, S., Lee, J. C., Ly, C., et al. 2014, \apj, 797, 126
\bibitem[\protect\citeauthoryear{S\'{a}nchez et al. 2013}{2013}]{San0} S\'{a}nchez, S. F., Rosales-Ortega, F. F., Jungwiert, B., et al. 2013, A\&A, 554, 58
\bibitem[\protect\citeauthoryear{S\'{a}nchez et al. 2017}{2017}]{San0} S\'{a}nchez, S. F., Barrera-Ballesteros, J. K., S\'{a}nchez-Menguiano, L., et al. 2017, \mnras, 469, 2121
\bibitem[\protect\citeauthoryear{Sanders et al. 2015}{2015}]{San1} Sanders, R. L., Shapley, A. E., Kriek, M., et al. 2015, \apj, 799, 138
\bibitem[\protect\citeauthoryear{Sanders et al. 2018}{2018}]{San2} Sanders, R. L., Shapley, A. E; Mariska, K., et al. 2018, \apj, 858, 99
\bibitem[\protect\citeauthoryear{Shen et al. 2003}{2003}]{Sa2} Shen, S., Mo, H. J., White, S. D. M., et al. 2003, /mnras, 343, 978
\bibitem[\protect\citeauthoryear{Shimasaku et al. 2001}{2001}]{Sa2} Shimasaku, K., Fukugita, M., Doi, M., et al. 2001, AJ, 122, 1238
\bibitem[\protect\citeauthoryear{Stott et al. 2013}{2013}]{St0} Stott, J. P., Sobral, D., Bower, R., et al. 2013, MNRAS, 436, 1130
\bibitem[\protect\citeauthoryear{Tremonti et al. 2004}{2004}]{Tr1} Tremonti, C. A., Heckman, T. M., Kauffmann, G., et al. 2004, \apj, 613, 898
\bibitem[\protect\citeauthoryear{Troncoso et al. 2014}{2014}]{Tr1} Troncoso, P., Maiolino, R., Sommariva, V. et al. 2014, A\&A, 563, 58
\bibitem[\protect\citeauthoryear{Wu et al. 2013}{2013}]{Wu1} Wu, Y.-Z., \& Zhang, S.-N. 2013, \mnras, 436, 934
\bibitem[\protect\citeauthoryear{Wu et al. 2016}{2016}]{Wu2} Wu, Y.-Z., Zhang, S.-N., Zhao, Y.-H., \& Zhang, W. 2016, \mnras, 457, 2929
\bibitem[\protect\citeauthoryear{Wuyts et al. 2014}{2014}]{Wuy0} Wuyts, E., Kurk, J., Forster Schreiber, N. M., et al. 2014, \apj, 789, 40
\bibitem[\protect\citeauthoryear{Wuyts et al. 2012}{2012}]{Wuy1} Wuyts, E., Rigby, J. R., Sharon, K., \& Gladders, M. D. 2012, \apj, 755, 73
\bibitem[\protect\citeauthoryear{Wuyts et al. 2016}{2016}]{Wuy2} Wuyts, E., Wisnioski, E., Fossati, M., et al. 2016, \apj, 827, 74
\bibitem[\protect\citeauthoryear{Yabe et al. 2015}{2015}]{Y0} Yabe, K., Ohta, K., Akiyama, M., et al. 2015, PASJ, 67, 102
\bibitem[\protect\citeauthoryear{Yabe et al. 2014}{2014}]{Y2} Yabe, K., Ohta, K., Iwamuro, F., et al. 2014, \mnras, 437, 3647
\bibitem[\protect\citeauthoryear{Yates et al. 2012}{2012}]{Y3} Yates, R. M., Kauffmann, G., \& Guo, Q. 2012, \mnras, 422, 215
\bibitem[\protect\citeauthoryear{Yates et al. 2014}{2014}]{Y4} Yates, R. M., \& Kauffmann, G. 2014, \mnras, 439, 3834
\bibitem[\protect\citeauthoryear{Zahid et al. 2012}{2012}]{Z4} Zahid, H. J., Bresolin, F., Kewley ,L. J., Coil, A. L., \& Dav\'{e}, R. 2012, \apj, 750, 120
\bibitem[\protect\citeauthoryear{Zahid et al. 2015}{2015}]{Z5} Zahid, H. J., Damjanov, I., Geller, M. J., \& Chilingarian, I. 2015, \apj, 806, 122
\bibitem[\protect\citeauthoryear{Zahid \& Geller 2017}{2017}]{Z3} Zahid, H. J \& Geller, M. J. 2017, \apj, 841, 32
\bibitem[\protect\citeauthoryear{Zahid et al. 2013}{2013}]{Z1} Zahid, H. J., Geller, M. J., Kewley, L. J., et al. 2013, \apjl, 771, 19
\bibitem[\protect\citeauthoryear{Zahid et al. 2014}{2014}]{Z2} Zahid, H. J., Kashino, D., Silverman, J. D., et al. 2014, \apj, 792, 75
\bibitem[\protect\citeauthoryear{Zaritsky et al. 1994}{1994}]{Z3} Zaritsky, D., Kennicutt, R. C., \& Huchra, J. P. 1994, \apj, 420, 87
\bibitem[\protect\citeauthoryear{Zhang et al. 2008}{2008}]{Z5}Zhang, W., Kong, X., \& Cheng, F. Z. 2008, CHJAA, 8, 211

\end{thebibliography}
\end{document}